 \pdfoutput=1
\documentclass{elsarticle}
\usepackage{rotating}

\usepackage{url}
\usepackage{amsmath}
\let\Phi\varPhi
\let\Theta\varTheta
\usepackage{array}

\journal{Future Generation Computer Systems}

\begin{document}

\begin{frontmatter}

\title{A Survey on Modeling Energy Consumption of Cloud Applications: Deconstruction, State of the Art, and Trade-off Debates}

\author{Zheng Li}
\address{Department of Electrical and Information Technology, Lund University, Sweden}
\ead{zheng.li@eit.lth.se}

\author{Selome Tesfatsion}
\address{Department of Computing Science, Ume{\aa} University, Ume{\aa}, Sweden}
\ead{selome@cs.umu.se}

\author{Saeed Bastani}
\address{Department of Electrical and Information Technology, Lund University, Sweden}
\ead{saeed.bastani@eit.lth.se}

\author{Ahmed Ali-Eldin}
\address{Department of Computing Science, Ume{\aa} University, Ume{\aa}, Sweden}
\ead{ahmeda@cs.umu.se}

\author{Rajiv Ranjan}
\address{School of Computer Science, Newcastle University, Newcastle Upon Tyne, UK}
\ead{raj.ranjan@ncl.ac.uk}

\author{Erik Elmroth}
\address{Department of Computing Science, Ume{\aa} University, Ume{\aa}, Sweden}
\ead{elmroth@cs.umu.se}

\author{Maria Kihl}
\address{Department of Electrical and Information Technology, Lund University, Sweden}
\ead{maria.kihl@eit.lth.se}

\begin{abstract}
Given the complexity and heterogeneity in Cloud computing scenarios, the modeling approach has widely been employed to investigate and analyze the energy consumption of Cloud applications, by abstracting real-world objects and processes that are difficult to observe or understand directly. It is clear that the abstraction sacrifices, and usually does not need, the complete reflection of the reality to be modeled. Consequently, current energy consumption models vary in terms of purposes, assumptions, application characteristics and environmental conditions, with possible overlaps between different research works. Therefore, it would be necessary and valuable to reveal the state-of-the-art of the existing modeling efforts, so as to weave different models together to facilitate comprehending and further investigating application energy consumption in the Cloud domain. By systematically selecting, assessing and synthesizing 76 relevant studies, we rationalized and organized over 30 energy consumption models with unified notations. To help investigate the existing models and facilitate future modeling work, we deconstructed the runtime execution and deployment environment of Cloud applications, and identified 18 environmental factors and 12 workload factors that would be influential on the energy consumption. In particular, there are complicated trade-offs and even debates when dealing with the combinational impacts of multiple factors. 
\end{abstract}

\begin{keyword}
Application energy consumption\sep Cloud computing\sep energy consumption modeling\sep energy-related factors\sep systematic literature review 
\end{keyword}

\end{frontmatter}


\section{Introduction}

Given the requirement of efficient use of computing power and the increasing consideration of global warming, the energy consumption management is a crucial concern across the entire community of the information and communication technology (ICT), especially in the Cloud computing domain \cite{Baliga_Ayre_2011}. In particular, understanding Cloud applications' energy consumption has been identified to be a prerequisite for developing energy saving mechanisms \cite{Feller_Ramakrishnana_2015}. Unfortunately, due to Cloud applications' inherent complexity and their environmental heterogeneity, it would be extremely challenging to tune the energy efficiency of a real-world application \cite{Barroso_Holzle_2007}, and even unpractical to directly measure its energy consumption. On one hand, the components and data of a modern application could largely be distributed and spread in Cloud environments. On the other hand, the same computing resource in the Cloud could be shared among a bunch of different applications. 

Consequently, most of the related work focused on the energy expense in the Cloud infrastructure and IT equipment (e.g., data center energy consumption \cite{Hammadi_Mhamdi_2014,Shuja_Bilal_2016}), without considering specific application scenarios or isolating a single application from its surroundings. In particular, with a lack of concern about the application runtime, some of the studies essentially emphasized the power consumption in Cloud systems from the hardware's perspective (e.g., \cite{Dayarathna_Wen_2016}). Note that here power (measured in \textit{Watts}) is defined as the rate at which energy (measured in \textit{Joules}) is consumed in the Cloud infrastructure.

As for the studies investigating Cloud applications' energy consumption, researchers tend to employ the modeling approach to relieve the aforementioned challenges and complexity, by abstracting real-world objects or processes that are difficult to observe or understand directly \cite{Reference_2016}. However, since such an abstraction sacrifices (and usually does not need) the complete reflection of the reality to be modeled, current energy consumption models vary in terms of purposes, assumptions, application characteristics and environmental conditions, with possible overlaps between different research works. As a result, different models need to be weaved together to reflect a full scope of energy consumption aspects, which is also common in other domains \cite{Mellor_Clark_2003}.

Therefore, to facilitate understanding the nature of the energy consumption of Cloud applications, it would be useful and valuable to come up with the state-of-the-art of the existing modeling efforts that play an evidence role in revealing the reality. When it comes to the evidence aggregation for answering research questions in software engineering and computer science, a standard and rigorous methodology is Systematic Literature Review (SLR) \cite{Kitchenham_Charters_2007}. Thus, we implemented an SLR to identify, examine and synthesize the existing models developed/employed in the relevant studies. Moreover, to help analyze and compare the existing models, we followed the divide-and-conquer strategy to also study the prerequisites of modeling practices: (1) Since the energy for running a Cloud application is driven by the combined mutual effects of the application and its environment \cite{Liu_Pinto_2015}, we extracted nine generic application execution elements and built up an evidence-based architecture of the application deployment environment. (2) Considering that Cloud computing scenarios involve numerous and various factors \cite{Miettinen_Nurminen_2010}, we identified 18 environmental factors and 12 workload factors respectively as well as their individual influences on Cloud applications' energy consumption.

Driven by the aforementioned motivations, our main contributions to the research field can be summarized as follows. First, our deconstruction of Cloud application runtime and deployment environment offers an expandable dictionary of energy-related factors. Benefiting from this dictionary, researchers and practitioners can conveniently screen the existing concerns and choose suitable ones for new energy consumption studies. In fact, pre-listing all the domain-relevant factors has been considered to be a ``tedious but crucial task" for factorial studies in general \cite{Boudec_2011,Li_OBrien_2012}. Second, the systematically organized models with unified notations can act as a knowledge artefact for both researchers and practitioners to not only reveal the fundamentals of energy consumption, but also facilitate simulations to deal with a wide range of Cloud application energy efficiency problems. For example, accurate model-based energy consumption simulations would be significantly beneficial for decision making in various trade-off situations. 

The remainder of this paper is organized as follows. Section \ref{sec:methodology} briefly describes the methodology employed in our survey, and particularly highlights the research questions and selection \& exclusion criteria. Section \ref{sec:slrResults} specifies the results of this survey by addressing the predefined research questions. Section \ref{sec:futuredirections} lists four trade-off debates to demonstrate both the complexity in combinational effects of multiple factors, and the potential research directions that can benefit from our survey. Conclusions and our future work are outlined in Section \ref{sec:conclusion}.




\section{Implementation Methodology of the Survey}
\label{sec:methodology}
Given the widely accepted SLR guidelines \cite{Kitchenham_Charters_2007}, we implemented our survey following a three-stage procedure, namely designing, conducting and reporting. Due to the space limit, we particularly highlight the research questions that essentially drive this literature review, and the inclusion and exclusion criteria that justify our study selection, while only briefly introducing our review conducting process together with the other details.
\subsection{Research Questions}

During the whole lifecycle of Cloud applications, energy consumption happens mainly when they are being deployed and executed \cite{Armstrong_Kavanagh_2015}. Moreover, as mentioned previously, the energy for executing a Cloud application is essentially caused by the combined mutual effects between the application software and its environmental infrastructure \cite{Liu_Pinto_2015}. Therefore, we decided to summarize the deployment environments and the runtime execution elements of Cloud applications:
\begin{description}
    \item[\textbf{RQ1}] What deployment environments of Cloud applications have been discussed in the relevant studies?
    \item[\textbf{RQ2}] What execution elements of Cloud applications have been discussed in the relevant studies?
\end{description}

Although there is no doubt that running Cloud applications will cause energy consumption, it is more valuable to identify influential factors to understand why different amounts of energy could be consumed even for the same application to achieve the same (or comparable) performance quality. Following the previous research questions, it is natural to distinguish between the environmental factors and the application workload factors: 
\begin{description}
    \item[\textbf{RQ3}] What environmental factors and their influences have been studied in Cloud application energy consumption?
    \item[\textbf{RQ4}] What workload factors and their influences have been studied in Cloud application energy consumption?
\end{description}

Through reviewing the modeling studies, one of our main purposes is to reveal Cloud applications' energy consumption models, because the mathematical models can theoretically explain how the energy is consumed:
\begin{description}
    \item[\textbf{RQ5}] What models have been developed for abstracting the energy consumption of Cloud applications?
\end{description}

\subsection{Inclusion and Exclusion Criteria}

In addition to the research questions, we also pre-clarify a set of inclusion and exclusion criteria to further shape our research scope, as specified below:

~~\\
\noindent
\textbf{\textit{Inclusion Criteria:}}
\begin{enumerate}
\renewcommand{\labelenumi}{\it{\theenumi)}}
    \item	Publications that profile/characterize the energy consumption of applications deployed in the Cloud environment.
    \item	Publications that investigate the energy consumption of local applications that have interactions with a Cloud system (e.g., workload offloading).
    \item	Publications that model application's (or application component's) energy consumption by denoting the energy consumption of environmental hardware.
    \item	Publications that reflect the changes in energy consumption of a Cloud-based application (or application component) by measuring hardware's energy consumptions with different workload configurations.
    \item	Publications that reflect the changes in energy consumption of a Cloud-based application (or application component) by measuring hardware's energy consumptions with different environmental configurations.
    \item	Publications that provide first-hand and relatively strong evidence through evaluations and peer reviews, such as book chapters and full journal/conference/workshop papers.
\end{enumerate}

\noindent
\textbf{\textit{Exclusion Criteria:}}
\begin{enumerate}
\renewcommand{\labelenumi}{\it{(\theenumi)}}
    \item	Publications that investigate the energy consumption of applications running in local environment (e.g., desktop systems) without addressing any concern related to the Cloud.
    \item	Publications that compare energy-saving strategies/algorithms through experiments without energy consumption modeling or factor discussions in a generic sense.
    \item	Publications that investigate the energy consumption of packet/frame transferring in the lower layers of network protocol stack (e.g., \cite{Xiao_Cui_2014}). Given our focus on the energy consumption in the application layer, we are concerned with bit/Byte/file data transmission.
    \item	Publications that investigate the energy consumption of a Cloud system or its components (e.g., server, cluster or datacenter \cite{Leverich_Kozyrakis_2010,Tian_Xiong_2013}) without regarding to a single application (component) scenario. In other words, this type of studies could be concerned with the overall workloads from numerous and various applications.
    \item	Publications that model the environmental hardware's energy consumption by notating applications' (or application components') energy consumption (e.g., \cite{Cappiello_Datre_2013}). This type of studies were not in the context of a single application (component) scenario, either.
    \item	Publications that do not contribute first-hand or strong evidence, such as survey papers (i.e.~secondary studies), extended abstracts, posters, short/position papers, and industry white papers.
\end{enumerate}

\subsection{Review Process}
By using the quasi-gold standard to manipulate search strings \cite{Zhang_Babar_2011}, we retrieved over 3000 publications from the five dominant electronic libraries (namely ACM Digital Library, Google Scholar, IEEE Xplore, ScienceDirect, and SpringerLink), and initially identified 394 studies through quickly scanning their titles and abstracts (note that we only screened the first 50 pages from Google Scholar). In particular, considering that the term ``Cloud computing" was coined in 2006 \cite{Zhang_Cheng_2010}, we did not search the literature published before 2006. 

After further examining the full texts of the initially collected studies against the inclusion \& exclusion criteria, we finally selected 76 papers to fit in this survey. It is notable that we have employed two strategies to reduce the selection bias and improve the fundamental reliability: Firstly, we conducted pilot reviews to try to well establish and polish the inclusion \& exclusion criteria in advance. Secondly, we organized regular meetings to discuss the unsure issues and cross-reviewed the borderline papers.

At last, a data extraction schema was developed to guide paper review and data identification in a structured fashion. In detail, the raw data were gradually extracted from the selected studies and aggregated into a big table to facilitate the overall data synthesis.\footnote{The schema together with the extracted raw data have been shared online: \url{https://goo.gl/JN8r7W}} Based on the data analysis, we deliver the review results and discussions by respectively addressing the aforementioned research questions, as specified in the following section.

\section{Review Results and Discussions}
\label{sec:slrResults}

\subsection{Deployment Environment of Cloud Applications (RQ1)}
\label{subsec:environment}
\begin{figure}[!t]
\centering
\includegraphics[width=11cm]{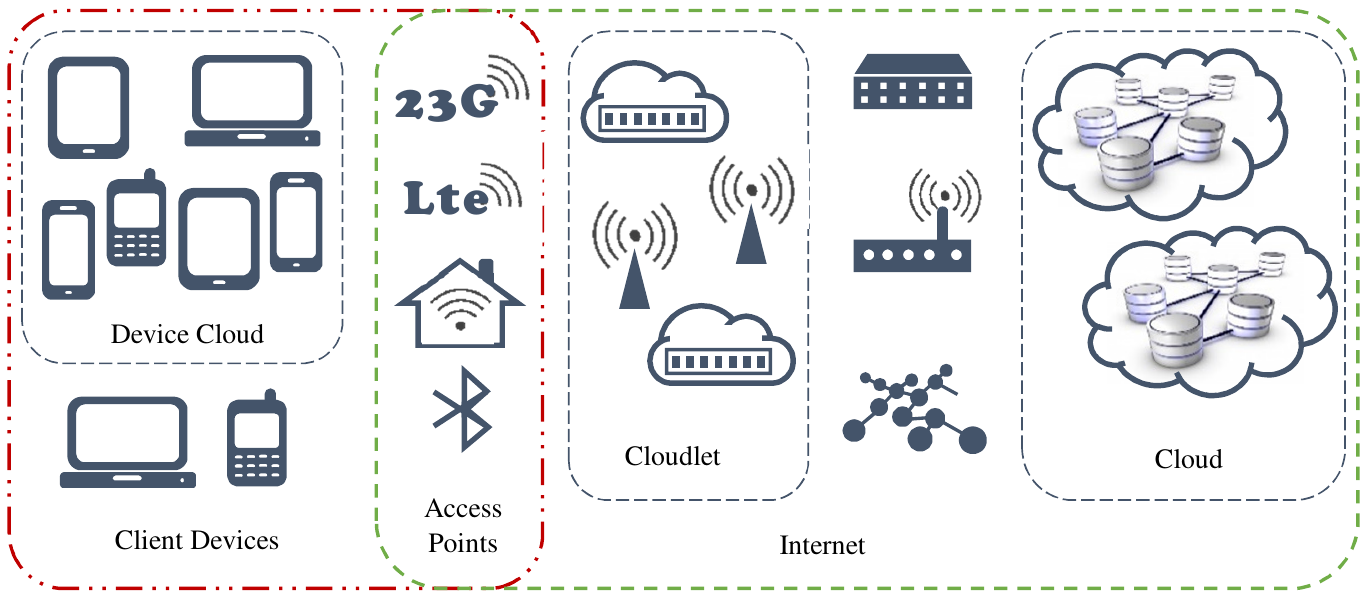}
\caption{Evidence-based environmental architecture for deploying Cloud applications.}
\label{PicEnvironment}
\end{figure}

It has been identified that the deployment environment has significant effects on the energy consumption of Cloud applications \cite{Gribaudo_Ho_2015}. Recall that a Cloud application is generally based on a multi-resource collaboration, and the application tasks could be deployed into different places typically including local devices and Cloud virtual machines \cite{Wu_Huang_2013}.
To facilitate locating the energy consumption sources when running Cloud applications, it would be useful to outline a generic deployment architecture in the context of Cloud computing. By extracting the information about deployment configurations from the reviewed studies, we draw an evidence-based environmental architecture for Cloud application deployment, as shown in Fig.~\ref{PicEnvironment}. 

\begin{itemize}

\item \textbf{Cloud:} Being located at the far end of the deployment architecture (cf.~Fig.~\ref{PicEnvironment}), the Cloud provides on-demand computing resources for users through the Internet. The Cloud computing paradigm is initially a business model by allowing Cloud consumers to avoid upfront infrastructure costs \cite{Li_Zhang_2013}. Driven by the requirement of energy efficiency in ICT, Cloud computing has acted as a promising solution to the global demand for green computing \cite{Altamimi_Naik_2011,Baliga_Ayre_2011,Ren_Zhang_2010}. Although the data centers in production could continuously use tremendous amounts of electricity \cite{Bonner_Namin_2012}, the Cloud has been advocated to be more environmentally friendly than local computing systems, for multiple reasons ranging from the improvement of utilization through resource multitenancy to the replacement of high-power local equipment with lightweight client devices \cite{Ali_2009,Liu_Wang_2009,Ren_Zhang_2010,Vishwanath_Jalali_2015,Yuan_Kuo_2010}. 

\item \textbf{Cloudlet:} The emergence of Cloudlet is a crucial evolution in mobile Cloud computing \cite{Altamimi_Naik_2011}. As the mobile and wearable devices are becoming pervasive, the mobile application market is booming \cite{Mtibaa_Harras_2013}. Many Cloud-based mobile applications require low latencies and high data throughput for their remote interactions and/or workload offloading. However, given the large separation between the local devices and the Cloud, moving computation tasks and transferring data have to go through WAN-scale network hops, which would consequently consume considerable energy and incur unacceptable delay and jitters \cite{Ma_Lin_2015}. To satisfy the resource and performance requirement of mobile applications, a natural approach is to push the Cloud closer to its end users. A Cloudlet can be viewed as a mobile-service-oriented and small-scale data center that is beside the clients or at the inner edge of the Internet. Some empirical studies have shown that, because of smaller round-trip delay, the nearby Cloudlet presents a better offloading option for computation-intensive workloads than the distant Cloud \cite{Mtibaa_Harras_2013,Ravi_Peddoju_2015}.

\item \textbf{Internet:} Recall that accessing the Cloud/Cloudlet relies on the de facto Internet infrastructure \cite{Altamimi_Naik_2011,Ravi_Peddoju_2015,Sheng_Mahapatra_2015}, and thus the Internet plays an irreplaceable role in the Cloud ecosystem. According to the telecommunication network design principles, the Internet can be segmented into three main parts including access, metro/edge, and core networks \cite{Hinton_Baliga_2011,Vishwanath_Jalali_2015}, besides the content distribution networks and data centers. Such a segment model has been used to estimate the overall power consumption in the Internet by integrating those individual components \cite{Baliga_Hinton_2007,Hinton_Baliga_2011}. From the application's perspective, however, the calculation of energy for data transportation through the Internet only comprises a small set of involved network equipment (cf.~Equation \ref{eqn:Internet} in Section \ref{subsubsec:communicationModel}). Therefore, to be aligned with the studies on Cloud applications' energy consumption, we simplify the Internet model to be an equipment combination of switches, routers and various links, plus the Cloudlet and Cloud, as illustrated in Fig.~\ref{PicEnvironment}.    

\item \textbf{Device Cloud:} Considering the potentially spare computing resources of surrounding devices, peer-device offloading has been proposed as an effective option to share workloads through Bluetooth ad-hoc network \cite{Ravi_Peddoju_2015}. A simulation-based theoretical analysis even showed 63\% more energy saving than traditional offloading to the Cloud \cite{Sheng_Hu_2014}. In addition to the cooperation between peer devices, the paradigm of device Cloud has naturally evolved from the increasing average quantity of mobile devices per user or household, for running an application among a set of cooperative devices \cite{Mtibaa_Harras_2013,Mtibaa_Snobery_2014,Song_Cui_2014}. By employing different wireless communication access technologies (e.g., WiFi, 2G/3G, LTE, etc.) and including sensors of various kinds (e.g., GPS, camera sensor, air pollution sensor, etc.), the cooperation among sensor nodes can be extended to a broad range, namely mobile wireless sensor network \cite{Sheng_Hu_2014}. As a matter of fact, the latest radio frequency technologies and enhanced processing capability make lightweight wireless sensor nodes also feasible to host sensing applications. Since a sensor is inevitably integrated into a particular electronic equipment (e.g., environmental monitor and vehicle diagnostic board) on the client side (or outer edge of the Internet \cite{Sheng_Mahapatra_2015}), we still treat the mobile wireless sensor network as part of the device Cloud paradigm.

\item \textbf{Client Device:} Although there are various types of client devices, the client-side energy consumption of Cloud applications has been discussed largely with respect to mobile handsets such as smartphones and tablets. In fact, mobile devices nowadays are becoming the primary computing platform and a mandatory part of daily life for many users \cite{Akram_ElNahas_2015,Deng_Huang_2015,Kumar_Lu_2010,Mtibaa_Harras_2013}. Unfortunately, due to the slow development of battery technology compared to the semiconductor technologies \cite{Altamimi_Naik_2011,Saab_Saab_2015}, the limited battery capacity has been identified to be a major bottleneck of mobile handsets, in contrast to the wall-socket-powered platforms   \cite{Chun_Ihm_2011,Kim_2012,Munoz_Iserte_2015}. Moreover, given the high demand for computationally expensive Cloud applications (e.g., the increasingly popular use cases of multimedia streaming), the client devices would further experience a significant increase in the local energy consumption \cite{Gao_Li_2014,Lin_Hsiu_2014,Segata_Bloessl_2014}. Correspondingly, the relevant studies are pervasively concerned with workload offloading strategies in mobile Cloud computing, in order to alleviate the suffering from the clients' energy shortage.

\end{itemize}

\subsection{Execution Elements of Cloud Applications (RQ2)}
\label{subsec:activities}
\begin{figure}[!t]
\centering
\includegraphics[width=10cm]{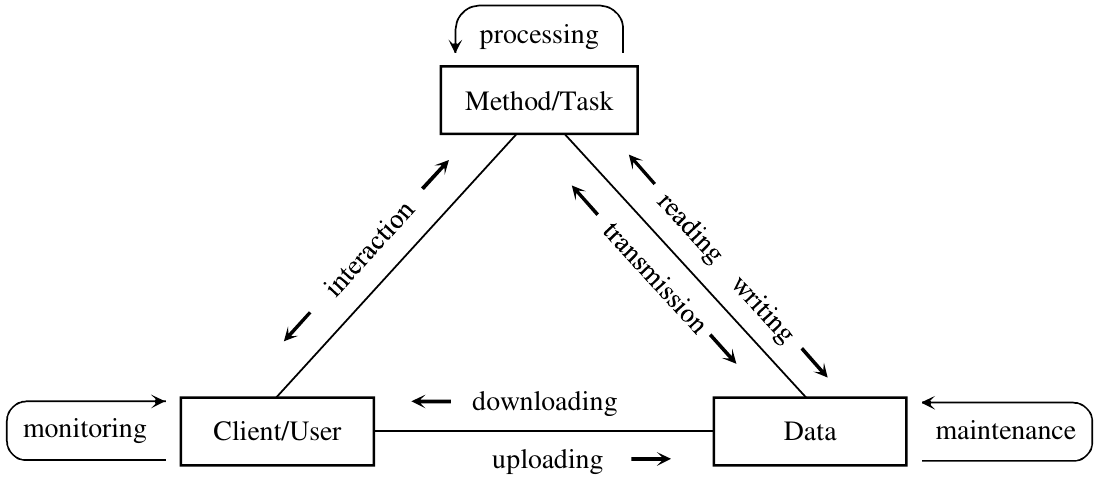}
\caption{Evidence-based organization of energy-relevant execution elements of Cloud applications.}
\label{PicActivities}
\end{figure}

Although there could be an infinite variety in functionality of Cloud applications, we emphasize generic execution elements. To facilitate identifying execution elements of Cloud applications, we pre-list three entities (namely Client/User, Method/Task, and Data) that drive, or are driven by, potential execution elements. 
At last, nine runtime elements across those entities are extracted from the reviewed papers, as shown in Fig.~\ref{PicActivities}. The discussion about application execution elements is combined into Section \ref{subsubsec:RQ2Discussion}. 

\begin{itemize}

\item \textbf{Downloading \& Uploading:} In essence, downloading/uploading indicates data access from the user's point of view. We recognize these two activities only when they are specifically discussed in the primary studies, for example the file uploading and downloading from the Cloud \cite{Baliga_Ayre_2011,Enokido_Suzuki_2010_AINA,Enokido_Suzuki_2010_CISIS}. In addition, since the radio frequency module (RF) of mobile devices demands different amounts of energy for sending and receiving data respectively (uploading generally costs more energy than downloading with respect to the same amount of data) \cite{Mtibaa_Harras_2013,Segata_Bloessl_2014}, we also employ this execution element to cover the separate uplink and downlink wireless transmissions \cite{Munoz_Iserte_2015,Xiang_Lin_2015}.

\item \textbf{Interaction:} Although the interaction between the client and the remote tasks essentially incurs data exchanging, investigating the energy consumption of interactive workloads could be particularly challenging, due to the fine granularity of communication \cite{Miettinen_Nurminen_2010}. Moreover, to intentionally study the mutual actions between a Cloud application and its users, it would be useful to distinguish interaction from the other types of communication elements. For example, instead of reflecting communication data throughput, this execution element is often highlighted when stressing the server load, like user connections \cite{Tang_Dai_2011} and user requests for playing online games \cite{Ravi_Peddoju_2015} or for exploring HTTP websites \cite{Chen_Grundy_2015,Murwantara_Bordbar_2014}. 

\item \textbf{Maintenance:} If a Cloud application requires data storage, one of its fundamental execution elements would be maintaining the availability and integrity of data. In practice, it is common to spread data across different locations to improve the data accessibility and reduce the likelihood of data loss \cite{Barroso_Holzle_2007}. Given the limited maintenance scenarios in the selected studies, we roughly identify data files to be stored either in the remote data centers (e.g., when employing storage as a service) or in the local client devices (e.g., when offloading computational workloads only) \cite{Baliga_Ayre_2011}. When it comes to the remote data maintenance, storing popular contents in the Cloudlet instead of the Cloud has widely been accepted as an energy-efficient strategy, for reducing the Internet traffic between the content data and their end users \cite{Altamimi_Naik_2011}.

\item \textbf{Monitoring:} When employing Cloud services, monitoring is one of the primary execution tasks especially in thin-client scenarios \cite{Baliga_Ayre_2011}. Considering the limited battery capacity of handset devices, runtime monitoring could be a major concern for energy consumption of mobile Cloud applications \cite{Chun_Ihm_2011}. Correspondingly, it has been proposed to scale the image frames' backlight levels in particular Cloud applications, like video streaming, in order to reduce the energy consumed in display modules of client devices \cite{Lin_Hsiu_2014}.

\item \textbf{Processing:} As the name suggests, we treat processing as the processor-centric execution element, such as mathematical calculation (e.g., generating a particular Fibonacci number \cite{Pu_Xu_2013}), logic task execution (e.g., workload-resource scheduling \cite{Chen_Schneider_2012,Luo_Yang_2015}), and data processing (e.g., mapping, shuffling and reducing the input data \cite{Feller_Ramakrishnana_2015}). Since processor has been considered to be the major power consumer in Cloud computing scenarios \cite{Lefurgy_Wang_2008}, processing seems to be the commonest energy-consuming activity that has been discussed in nearly all the selected studies.   

\item \textbf{Reading \& Writing:} Compared to data accessing from the user's point of view (i.e.~Downloading \& Uploading), the application task's perspective considers two types of energy consumption elements of data accessing. The first type focuses on data reading/writing from/to where the data are stored, while the second type emphasizes data transmission through the network. Although not specified in every study, these two element types usually coexist with each other in Cloud applications (e.g., the data fetching requires both disk reading and network transferring \cite{Izadpanah_Pawlikowski_2013}). When it comes to Reading \& Writing only, one trend is that disk IO is more power-consuming than memory IO, while another trend is that data writing is generally more power-expensive than reading \cite{Liu_Pinto_2015}. 

\item \textbf{Transmission:} As mentioned above, the element data transmission mainly focuses on application tasks with respect to their data transfer over network resources. Since different tasks of a Cloud application can be executed distributedly, the data transmission could take place not only in the Cloud but also between the Cloud and the client (note that we identify Cloud-client data transmission from a study when it does not emphasize Downloading \& Uploading or Interaction). In either case, a Cloud application that transfers large amounts of data would cause a significant proportion of its whole energy consumption, due to two facts: (1) In the Cloud, routers, switches, links and aggregation resources consume more than 30\% of the total energy  \cite{Kliazovich_Bouvry_2012}; (2) On the client side, data communication has significant impacts on mobile devices' energy consumption  \cite{Folino_Pisani_2014}.  

\end{itemize}

\subsubsection{Summary}
\label{subsubsec:RQ2Discussion}
\begin{figure}[!t]
\centering
\includegraphics[width=11.5cm]{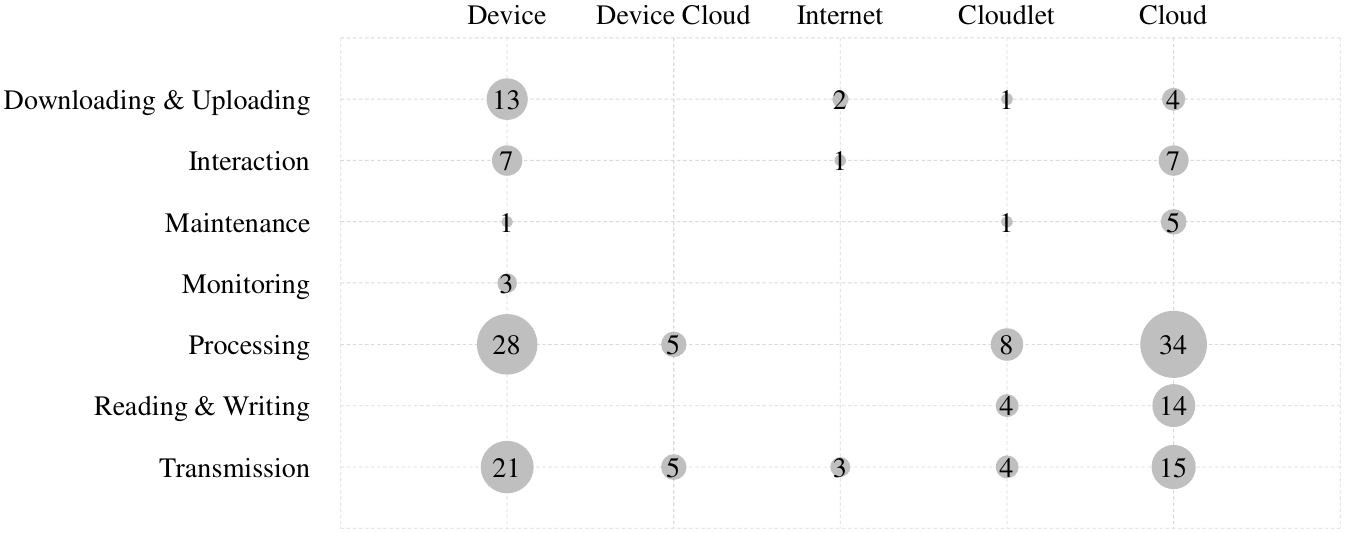}
\caption{Distribution of the energy consumption studies with respect to execution elements and deployment environment of Cloud applications.}
\label{PicBubble}
\end{figure}

According to the investigated execution elements and deployment environment of Cloud applications, we distribute the selected studies over a bubble plot, as shown in Fig.~\ref{PicBubble}. It is notable that the same study could have been counted in different bubbles, because one energy investigation might include multiple execution elements and different environmental components (e.g., \cite{Baliga_Ayre_2011}). With regarding to the execution elements, a clear trend is that most studies have focused on task processing and data transmission, which confirms computation and communication as two major concerns about a Cloud application's energy expense (e.g., using a communication-computation ratio to characterize application workloads and analyze its influence on energy efficiency \cite{Hsu_Lin_2013}). Among the environmental components for Cloud application deployment, client devices and Cloud have attracted the most research attentions. By examining their research methods, the reason seems to be twofold: (1) Client devices can directly be controlled and measured; and (2) Cloud data centers can be simplified into a local-server simulation, while the local servers are controllable and measurable. Such a distribution confirms that uncontrollable deployment environment makes addressing a Cloud application's energy consumption more challenging and complex. Correspondingly, by abstracting the uncontrollable aspects, modeling and model-based simulations would be a practical and effective research approach in this case. 

\subsection{Environmental Factors and their Influences on Energy Consumption of Cloud Applications (RQ3)}
\label{subsec:RQ4-2}
\begin{figure}[!t]
\centering
\includegraphics[width=11cm]{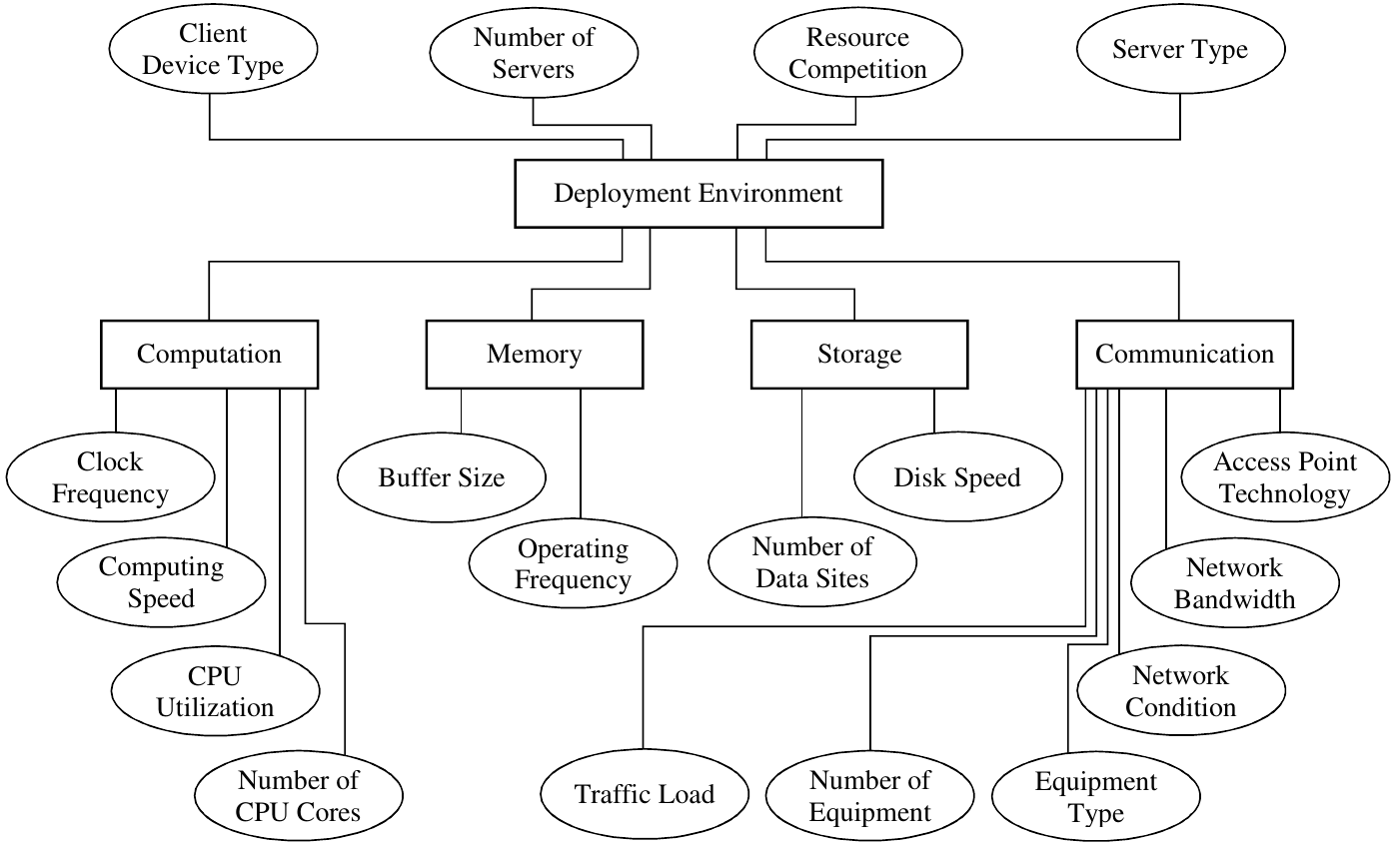}
\caption{Evidence-based deconstruction of environmental configurations of Cloud application deployment.}
\label{PicResources}
\end{figure}

Although the environmental architecture is straightforward (cf.~Section \ref{subsec:environment}), the deployment of a Cloud application could require sophisticated environmental configurations, and different environmental conditions might in turn drive different deployment strategies (e.g., the right data distribution with excellent connectivity would be wrong under poor communication channels \cite{Chun_Ihm_2011}). In essence, it is the detailed configurations that expose significant environmental impacts on the energy consumption of Cloud applications \cite{Chen_Grundy_2013,Chen_Grundy_2015}. To alleviate the complexity in energy analysis with various deployment configurations, it would be valuable to identify individual environmental factors and distinguish their energy influences between each other. Given the fine-grained decomposition of the IT infrastructure \cite{Li_OBrien_2014}, the existing studies were mainly concerned with four Cloud resource types, i.e.~computation, communication, memory and storage. We accordingly group and report the identified environmental factors, as organized through an entry-relationship diagram in Fig.~\ref{PicResources}. 

\subsubsection{Communication Environmental Factors} 
	\begin{enumerate}
	\renewcommand{\labelenumi}{\it{\theenumi)}}
		 \item	\textbf{Access Point Technology:} Nowadays diverse network technologies are available in different situations for accessing Cloud services, ranging from traditional Ethernet to modern cellular telecommunication. The energy consumption influenced by different technologies is mainly discussed with regarding to client devices \cite{Jalali_Gray_2014,Saab_Chehab_2013}. Among the popular access point technologies, WiFi and Ethernet generally consume less energy than cellular wireless networks \cite{Altamimi_Naik_2011,Chun_Ihm_2011,Fekete_Csorba_2012,Saab_Saab_2015,Vishwanath_Jalali_2015,Xia_Liang_2014}; although providing lower data rate, Bluetooth could be 80\% to 120\% more energy efficient than WiFi \cite{Mtibaa_Harras_2013}; as for the cellular networks, LTE (4G) consumes more power than UMTS (3G), followed by EDGE (2G) \cite{Ravi_Peddoju_2015,Segata_Bloessl_2014}.

		 \item	\textbf{Network Bandwidth:} As indicating the maximum channel capacity, the network bandwidth is considered to have a positive impact on reducing both the transmission delay and the energy consumption of Cloud applications \cite{Luo_Yang_2015,Yuan_Guo_2015}. Consequently, bandwidth has become a critical concern for computational offloading in the context of mobile Cloud computing \cite{Wu_Wolter_2015}: The offloading effort is not preferred until the connection has sufficient bandwidth, and the benefit of offloading enlarges as the network bandwidth increases \cite{Kumar_Lu_2010,Xia_Ding_2014}. In particular, in addition to the TCP stream bandwidth between different computing resources \cite{Balakrishnan_Tham_2013,Fard_Prodan_2012}, the researchers are also concerned with the bandwidth of network equipment (e.g., access points \cite{Mazza_Tarchi_2014,Xia_Liang_2014} and base station \cite{Deng_Huang_2015}).

   	 \item	\textbf{Network Condition:} Given the same communication coefficients, better channel quality improves Cloud applications' energy performance \cite{Sheng_Mahapatra_2015}, while poor network conditions worsens both response time and energy efficiency \cite{Akram_ElNahas_2015,Nabi_Mittal_2015}. The network condition can be reflected by the signal strength or the signal to noise ratio \cite{Mazza_Tarchi_2014}. When the signal strength is low, the relevant network devices will have to increase their power levels for data transmission \cite{Ravi_Peddoju_2015}, and will correspondingly end up with higher communication cost \cite{Saab_Saab_2015}. Furthermore, weak signals would lead to high chance of network unavailability \cite{Wu_Huang_2013}. In the worst case, significant energy would be consumed for frequently reestablishing the broken connections, rather than actual data transmission  \cite{Gao_Li_2014}.  

		 \item	\textbf{Network Equipment Type:} Recall that the Internet topology involves various network equipment, while different types of equipment have different power profiles. Thus, the network equipment types are specified particularly when analyzing the communication energy consumption in Cloud applications \cite{Jalali_Gray_2014,Luo_Yang_2015}. For example, the energy for delivering one bit data through the Internet would be associated with the power consumed in multiple gateways, switches, routers, and high-capacity wavelength division multiplexed fiber links located in different network segments \cite{Altamimi_Naik_2011,Baliga_Ayre_2011,Vishwanath_Jalali_2015}.

		 \item	\textbf{Number of Network Equipment:} As mentioned above, a communication line could comprise multiple groups of identical network equipment, and in practice the data traversal would hop through different types of equipment at different amounts \cite{Altamimi_Naik_2011,Baliga_Ayre_2011}. In particular, the number of routers (and their power profiles) was emphasized for the energy expenditure along a data transmission path \cite{Izadpanah_Pawlikowski_2013}.

		 \item	\textbf{Traffic Load:} Although a network equipment's power profile is predefined by its manufacturer, its practical power consumption would vary depending on the equipment's traffic load \cite{Izadpanah_Pawlikowski_2013}. Meanwhile, the traffic load ratio also indicates the resource utilization level of network devices \cite{Luo_Yang_2015}. Similar to the CPU utilization, higher traffic load would increase the communication energy consumption for Cloud applications.

	\end{enumerate}

\subsubsection{Computation Environmental Factors} 
	\begin{enumerate}
	\renewcommand{\labelenumi}{\it{\theenumi)}}
		 \item	\textbf{Clock Frequency (and Supply Voltage):} CPU's power consumption is dominantly influenced by its supply voltage \cite{Wu_Wu_2015}. Since the supply voltage is about linearly proportional to the operating clock frequency \cite{Luo_Yang_2015}, and only frequency can be altered without making physical adjustments \cite{Bonner_Namin_2012}, most researchers have mainly focused on the clock frequency as a factor \cite{Babukarthik_Raju_2012,Ge_Feng_2010_IPDPSW,Hsu_Lin_2013,Kim_Beloglazov_2009,Ou_Pang_2012,Song_Cui_2014,Thanavanich_Uthayopas_2013}. Intuitively, scheduling low clock frequency will scale down the supply voltage, which eventually brings power saving for CPU \cite{Abdelsalam_Maly_2009}. With relax application deadlines, the frequency (or voltage) downscaling has become a preferable approach to energy saving \cite{Balakrishnan_Tham_2013,Sheng_Hu_2014,Sheng_Mahapatra_2015}, especially for non-CPU intensive workloads \cite{Ge_Feng_2010_TPDS,Ibrahim_Moise_2014,Liu_Pinto_2015,Wang_Laszewski_2010,Wirtz_Ge_2011,Zhang_Fu_2011}. In particular, fine-grained frequency levels seem to be more energy friendly for Cloud applications \cite{Kim_Buyya_2007,Huang_Su_2012}.

		 \item	\textbf{Computing Speed:} The capacity of a Cloud computational resource can be measured by its computing speed in millions of instructions per second (MIPS) \cite{Fard_Prodan_2012}. In general, maintaining high processing speed would consume more energy \cite{Sheng_Mahapatra_2015}. In mobile Cloud computing, the speeds of client devices and Cloud servers are usually discussed together, in order to calculate their computing speedup (i.e.~the Cloud-client computing speed ratio) \cite{Mazza_Tarchi_2014,Wu_Wolter_2015}. The bigger speedup might indicate the better offloading opportunity, and lead to the higher application performance and the lower energy consumption \cite{Kumar_Lu_2010,Deng_Huang_2015,Yuan_Guo_2015}.

		 \item	\textbf{CPU Utilization:} The studies \cite{Tang_Dai_2011,Vishwanath_Jalali_2015} considered the power consumption in a server to be an exponential function of its CPU utilization, and the high CPU utilization is related to the underlying large workload size. Accordingly, higher utilization would result in more energy consumption within the same size of time window \cite{Zhang_Fu_2011}. 

		 \item	\textbf{Number of CPU Cores:} The power consumption of a Cloud computational resource depends on the number of its active CPU cores \cite{Enokido_Takizawa_2015}, with a proportional linear relationship \cite{Bergen_Desmarais_2014}. When the physical cores are saturated, adding more workload will not further increase the resource's power usage \cite{Zhang_Fu_2011}. On the other hand, employing more CPU cores to address the increasing workload will significantly consume more energy due to the increased CPU power and parallelization overhead \cite{Wirtz_Ge_2011}. Thus, allocating more than enough resources will inevitably result in wastes of energy \cite{Bergen_Desmarais_2014}. Note that utilizing more computational resources to improve a Cloud application's processing concurrency is not a concern here. Multiple factors' combinational impact on energy consumption is discussed in Section \ref{sec:futuredirections}.

	\end{enumerate}

\subsubsection{Memory Environmental Factors}
	\begin{enumerate}
	\renewcommand{\labelenumi}{\it{\theenumi)}}
		 \item	\textbf{Buffer Size:} As a generally predefined factor, memory buffer size could have to be decided by developers before the Cloud application deployment. The experiments showed that buffering different sizes of data would be sensitively influential on the energy costs of not only the data I/O methods but also the data compression/decompression \cite{Ge_Feng_2010_IPDPSW,Liu_Pinto_2015,Singh_Naik_2015}. For file operations, buffer size between 64KB and 256KB seems to be the most energy-efficient setting \cite{Singh_Naik_2015}. 

		 \item	\textbf{Operating Frequency:} Memory operating frequency has been viewed as one of the fundamental contributors to the power consumption in memory \cite{Zhang_Fu_2011}. Similar to the CPU clock frequency, higher memory frequency will also consume more power.
	\end{enumerate}

\subsubsection{Storage Environmental Factors}
	\begin{enumerate}
	\renewcommand{\labelenumi}{\it{\theenumi)}}
		 \item	\textbf{Disk Speed:} Among all the indexes of a storage device, the disk speed is emphasized in the energy expenditure of an application's storage I/O operations. \cite{Izadpanah_Pawlikowski_2013}. The power characteristics of disk speed and other indexes are essentially determined by storage device manufacturers.


		 \item	\textbf{Number of Data Sites:} Spreading data across different sites is a common practice to improve data availability. Correspondingly, for a Cloud application, the more sites need to be visited, the more energy and time will be consumed for more data transmissions \cite{Izadpanah_Pawlikowski_2013}.

	\end{enumerate}

\subsubsection{Other Environmental Factors}
	\begin{enumerate}
	\renewcommand{\labelenumi}{\it{\theenumi)}}
		 \item	\textbf{Client Device Type:} Although various user handsets do not show big difference in energy consumption for running mobile Cloud applications \cite{Segata_Bloessl_2014}, the client device type indeed matters when making comparison among desktops, laptops and cell phones \cite{Jalali_Gray_2014,Saab_Chehab_2013,Song_Cui_2014}. Given different power profiles, replacing a personal computer with a low-power consuming device would make the same Cloud application more energy-efficient in a generic sense \cite{Vishwanath_Jalali_2015}. If emphasizing the overall share of power consumed in the device communication (e.g., the WiFi interface has a bigger share of the power consumption in smartphones than laptops), however, larger client devices seem preferable for Cloud applications with respect to their energy consumption \cite{Namboodiri_Ghose_2013}.  

		 \item	\textbf{Number of Servers:} In a Cloud host, provisioning more virtual machines could require more physical servers \cite{Zhang_Fu_2011}, and activating more physical servers implies enhancing the needed power level \cite{Hsu_Lin_2013}. Meanwhile, the increased maintenance overhead after provisioning more virtual machines will eventually increase the energy consumption per task in an application \cite{Chen_Grundy_2013}. Therefore, selecting a suitable number of servers should optimize the overall power consumption and the total workload \cite{Abdelsalam_Maly_2009}. Similar to the aforementioned factor of number of CPU cores, allocating more than enough servers will cause energy waste during the execution of a Cloud application, even if employing sophisticated energy saving mechanisms \cite{Zheng_Huang_2014}.

		 \item	\textbf{Resource Competition:} If holding the computing resource constant, fierce resource competition could dramatically increase the corresponding energy consumption, no matter what the resource (component) is. For example, configuring more virtual machines within the same physical server will increase the CPU activities and incur extra scheduling overhead \cite{Chen_Schneider_2012}. Hosting multiple application instances in a single virtual machine would consume more energy than running application instances separately  \cite{Gribaudo_Ho_2015}. As for the resource components, the intense competition for access point connections \cite{Mazza_Tarchi_2014}, CPU processes \cite{Fekete_Csorba_2012}, memory footprints \cite{Ge_Feng_2010_TPDS}, and disk IO bandwidth \cite{Zhang_Fu_2011} have all been proved negatively impacting Cloud applications' energy efficiency.

		 \item	\textbf{Server Type:} The relevant studies have addressed the types of physical server, virtual server and Web server for their influences on Cloud applications' energy consumption. The physical server type can further be defined by using processor number or types (e.g., Intel vs. ARM-based processors) \cite{Enokido_Suzuki_2010_AINA,Ou_Pang_2012}. Given a particular Cloud server pool, the large heterogeneity in server types will result in high variance in the application execution time \cite{Hsu_Lin_2013}. As for virtual servers, vertical scaling (adjusting the server type) has clear impacts on the energy consumption and performance of a Cloud application. However, the specific influences of different virtual machine types are closely related to the application types (workload characteristics) \cite{Chen_Grundy_2013}. For example, among different HTTP Web servers, Apache and Lighttp are more energy efficient for lightweight workload, while Nginx consumes relatively less power at faster user arrival speed \cite{Murwantara_Bordbar_2014}.

	\end{enumerate}

\subsubsection{Summary}
\label{subsubsec:EnvironFactorDicussion}
\begin{figure}[!t]
\centering
\includegraphics[width=11.5cm]{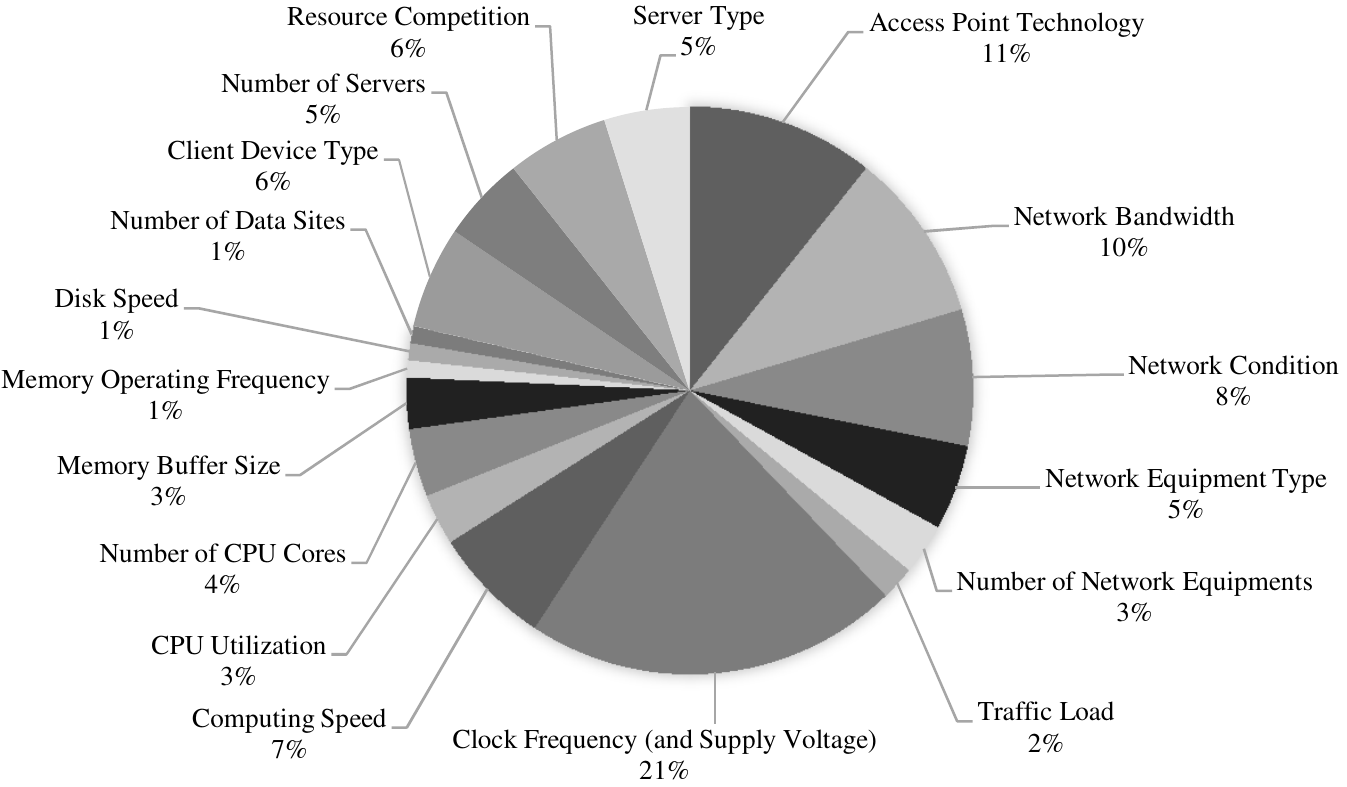}
\caption{Study popularity of environmental factors that are influential on energy consumption of Cloud applications. (The total number of factor-studies is 103.)}
\label{PicEnvironmentalFactors}
\end{figure}

 Overall, we have identified 18 environmental factors from the relevant studies. 
To facilitate tracing back to the reviewed studies, relevant publications are specified for each of the factors. Since the identified factors were not evenly studied, it would be useful to reveal to what extent those factors concerned researchers.
Here we employ factor-studies as a metric to measure the popularity of the identified factors, i.e.~one factor-study of a particular factor indicates that the factor is involved in one study. The popularity distribution is illustrated in Fig.~\ref{PicEnvironmentalFactors}.

It is clear that the CPU clock frequency has been studied as an outstanding environmental factor, followed by the technology of access points and the network bandwidth. As for the factor-study distribution over the four resource types, we only found five studies for two memory factors and one study for two storage factors. This huge imbalance in factor-studies further confirms computation and communication as two major energy concerns in the existing research work from the environmental perspective.

In particular, there are conflict opinions about adjusting CPU clock frequencies for energy saving, particularly through dynamic voltage and frequency scaling (DVFS). Although intelligently scaling frequency can improve energy efficiency, its benefits seem to be trivial \cite{Song_Cui_2014}, and the achievable energy saving could be 13\% \cite{Ge_Feng_2010_IPDPSW} to 20\% only \cite{Ou_Pang_2012}. Furthermore, different applications might have their best energy efficiency at different optimal frequencies \cite{Wirtz_Ge_2011}, and thus the same DVFS scheduling could only be sub-optimal for those different applications \cite{Ibrahim_Moise_2014}.

It is also notable for Access Point Technology that, although WiFi is generally more energy efficient than the cellular technologies, the superiority of WiFi becomes marginal if the utilization of cellular is high (for example when transmitting large bulks of data) \cite{Huang_Qian_2012}. Meanwhile, the efficiency of WiFi in saturation traffic would significantly degrade due to packet loss and retransmissions. 



\subsection{Workload Factors and their Influences on Energy Consumption of Cloud Applications (RQ4)}
\label{subsec:RQ4-1}
\begin{figure}[!t]
\centering
\includegraphics[width=11cm]{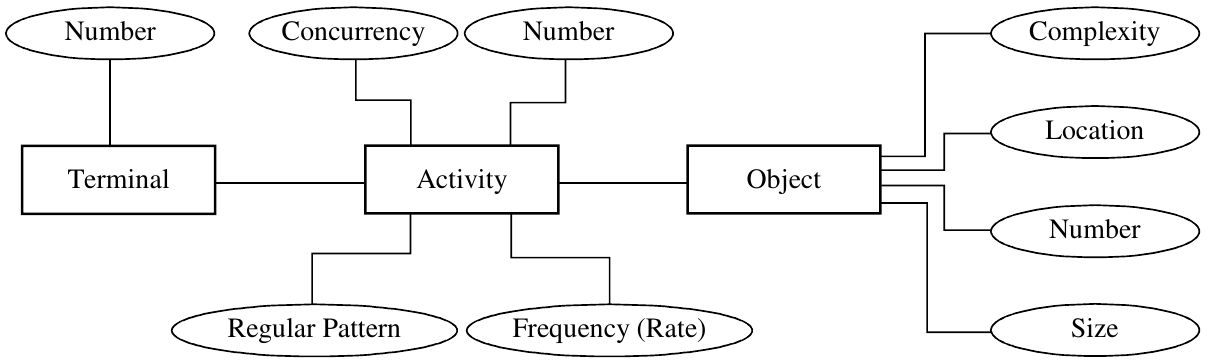}
\caption{Evidence-based deconstruction of Cloud application workloads.}
\label{PicWorkload}
\end{figure}

Since the energy for running a Cloud application is tightly coupled with its workload \cite{Chen_Schneider_2012,Luo_Yang_2015}, we identify energy-related factors by deconstructing Cloud application workloads. 
In Cloud environments, an application's workload can be described through one of three different aspects (namely Terminal, Activity, and Object) or a combination of them \cite{Li_OBrien_2014}. Correspondingly, we further organize the workload factors into those three aspects respectively, and use an entry-relationship diagram to illustrate the organization, as shown in Fig.~\ref{PicWorkload}. In particular, we consider application type to be an inherent attribute of a Cloud application, and thus ``application type" \cite{Abdelsalam_Maly_2009,Folino_Pisani_2014,Namboodiri_Ghose_2013,Thanavanich_Uthayopas_2013} is not regarded as a factor in our survey. In other words, we claim that the type of a Cloud application has already been reflected by its workload characteristics (e.g.~the specific communication-computation ratio).

\subsubsection{Terminal-related Factors} 
The client-side terminals usually act as workload generators in interaction-intensive Cloud applications. 
	\begin{enumerate}
	\renewcommand{\labelenumi}{\it{\theenumi)}}
    	 \item	\textbf{Number of Clients:} As workload generators, the client-side terminals can be either end users \cite{Vishwanath_Jalali_2015} or machines \cite{Enokido_Suzuki_2010_CISIS,Enokido_Suzuki_2010_AINA}, and the number of clients have been used to reflect the size of the generated workload. Naturally, the more number of clients an application serves, the more electric energy the application consumes.
	\end{enumerate}

\subsubsection{Activity-related Factors} 
Revoking the previous analysis in Section \ref{subsec:activities}, here we identify factors mainly related to those generic application execution elements.
	\begin{enumerate}
	\renewcommand{\labelenumi}{\it{\theenumi)}}
		\item 	\textbf{(Data) Access Pattern:} Data accessing refers to the reading and writing activities. Simple access patterns are relevant to activities only, such as one-time access, repeat access, and cyclic access; while sophisticated access patterns are associated with both the activities and the spatial distance between data locations, such as sequential access, nested access, and random access \cite{Ge_Feng_2010_IPDPSW}. When accessing the same amount of data, longer distance traversals will apparently consume more energy. For example, random access has been empirically verified to be significantly more energy expensive \cite{Liu_Pinto_2015}.

    	 \item	\textbf{(Data) Transmission Rate:} Without exceeding physical bandwidths, the power consumed in both servers and network equipment is a proportional function of the total data transmission rate in a Cloud application \cite{Enokido_Suzuki_2010_AINA,Enokido_Suzuki_2010_CISIS,Vishwanath_Jalali_2015}. However, data transfer at higher bit rate would be more energy efficient (i.e.~less energy consumption per bit) \cite{Miettinen_Nurminen_2010}, and therefore the downloading speed should be set as high as possible to save energy for client devices \cite{Munoz_Iserte_2015}. On the contrary, the energy consumption per bit was identified to be an increasing function of the data uploading rate from mobile devices. Considering that the low-speed traffic flows's impact on the overall power consumption is generally negligible \cite{Vishwanath_Jalali_2015}, decreasing the uploading speed has been argued to be an energy optimal solution on the client side (with flexible time limit) \cite{Munoz_Iserte_2015}.


    	 \item	\textbf{Number of (User) Connections:} For a Cloud application at runtime, one ``connection" indicates an active user session, no matter what activity is issued from the client side. When more user sessions are active, more energy consumption of the application will be incurred \cite{Chen_Grundy_2013,Tang_Dai_2011}. The user connections can be sequential, overlapped, or concurrent (e.g., file downloading from the Cloud \cite{Baliga_Ayre_2011}). In the concurrent case, more user activities would lead to an increase in Cloud resource usage, and the extra scheduling and synchronizing overhead could in turn increase each user request's processing time \cite{Chen_Grundy_2015}. 

		 \item	\textbf{Processing Concurrency:} Concurrent processing activities commonly exist in parallel applications, and the concurrency can be measured by the amount of processes. Due to the overhead of scheduling, both overall and per-task energy consumption could increase with the number of processes \cite{Chen_Schneider_2012,Chen_Grundy_2013}. However, unlike the other types of activities, the concurrency is generally for speeding up workload processing, rather than influencing the workload size. Accordingly, although incurring extra scheduling, increasing the degree of parallelism in a Cloud application can still significantly improve its energy efficiency (i.e.~the workload-energy ratio) \cite{Ge_Feng_2010_TPDS,Ou_Pang_2012,Wirtz_Ge_2011}. In particular, when memory footprints are relatively small, starting multiple processes within less computing resources can be even more energy friendly \cite{Bergen_Desmarais_2014}, until reaching the maximum utilization or physical limits of the resources (e.g., the total number of hyperthreads) \cite{Enokido_Takizawa_2015,Zhang_Fu_2011}.

    	 \item	\textbf{(User/Task) Arrival Rate:} Following the convention of the primary studies, we also use ``arrival rate" to represent the frequency of user interactions and task processing. In general, the faster user arrival rate \cite{Murwantara_Bordbar_2014} and the shorter inter-arrival time between two consecutive tasks \cite{Kim_Buyya_2007} both imply the tenser workload, and correspondingly result in the higher power consumption of a Cloud application. Note that the actual energy consumption eventually depends on the application's execution time, as specified above.

	\end{enumerate}

\subsubsection{Object-related Factors} 
Objects connect, and usually act as targets of, activities in workloads. Similarly, given our previous analysis in Section \ref{subsec:activities}, we identify data and task as two types of objects in Cloud applications. In particular, following the object-oriented thinking, a task can be viewed as a composite object (or a dividable piece of workload) that might include other types of workload elements.
	\begin{enumerate}
	\renewcommand{\labelenumi}{\it{\theenumi)}}
   	 \item	\textbf{Data Location:} Locality could be a significant contributor to the energy consumption of data accessing. As mentioned in \textbf{Data Access Pattern}, it is the data location that essentially impacts different patterns' influences \cite{Ge_Feng_2010_IPDPSW}. Thus, moving data closer to where they are needed seems to be an energy saving principle. For example, the collocated data and compute configuration delivers the best energy profile \cite{Saab_Saab_2015}, while distributing data and compute nodes into different layers will result in more energy consumption \cite{Feller_Ramakrishnana_2015}.

   	 \item	\textbf{Overall Data Size:} The existing studies exhibit a consensus on the positive correlation between the overall data size and the energy consumption of a Cloud application, even though the correlation was studied in various contexts. For instance, the input data size is a major driver behind the computation workload \cite{Sheng_Hu_2014,Song_Cui_2014}; the energy incurred by accessing activities mainly depends on the data length \cite{Izadpanah_Pawlikowski_2013}; and the amount of data to be transmitted is one of the discriminating factors for communication energy cost \cite{Folino_Pisani_2014,Kim_2012,Luo_Yang_2015}. In the context of communication, the relevant studies further distinguish between two scenes: The first is on the traffic volumes exchanged between the client and the Cloud \cite{Akram_ElNahas_2015,Baliga_Ayre_2011,Enokido_Suzuki_2010_AINA,Enokido_Suzuki_2010_CISIS,Fekete_Csorba_2012,Jalali_Gray_2014,Kumar_Lu_2010,Mazza_Tarchi_2014,Mtibaa_Harras_2013,Ou_Pang_2012,Saab_Chehab_2013,Saab_Saab_2015,Segata_Bloessl_2014,Vishwanath_Jalali_2015,Yuan_Guo_2015}, while the second is on the data segments involved in, and transferred between, individual application tasks \cite{Balakrishnan_Tham_2013,Chen_Schneider_2012,Chen_Grundy_2013,Deng_Huang_2015,Fard_Prodan_2012,Feller_Ramakrishnana_2015,Xia_Liang_2014}.

   	 \item	\textbf{Transactional Data Size:} Although the required energy increases proportionally to the overall data size, small-data transactions in a Cloud application show a negative correlation with the energy consumption. In practice, the data block per transaction can vary from several bytes to multiple megabytes \cite{Ge_Feng_2010_IPDPSW}. Given the same amount of data in an application, dealing with smaller-data-size transactions would cause longer execution time and higher energy expense \cite{Chen_Grundy_2013}. Consequently, packing a set of small data requests into a bulk transaction becomes an effective approach to improve the application's energy efficiency \cite{Ge_Feng_2010_IPDPSW,Miettinen_Nurminen_2010}. Note that the aforementioned data segments involved in application tasks do not necessarily act as transactional data pieces, because a task might further comprise numerous transactions.

    	 \item	\textbf{Number of Tasks:} By representing Cloud applications as task interaction graphs (e.g., directed acyclic graph), the number of task nodes and edges has been used to reflect the whole workload (i.e.~graph size) \cite{Balakrishnan_Tham_2013,Hsu_Lin_2013,Huang_Su_2012,Kim_Buyya_2007,Wu_Huang_2013,Zheng_Huang_2014}. Since more tasks usually imply more data and more application activities at runtime, the corresponding application execution will inevitably require more energy \cite{Izadpanah_Pawlikowski_2013}. Moreover, considering the extra overhead and energy for task scheduling, a larger number of tasks in a Cloud application will lead to higher average energy consumption per task \cite{Chen_Grundy_2013}.

    	 \item	\textbf{Task Complexity:} The computational complexity in tasks or functional modules is closely associated with the Cloud application's energy consumption \cite{Fekete_Csorba_2012,Liu_Cao_2014,Vishwanath_Jalali_2015}, as complex computation requires more computing resources and/or causes longer execution time. To verify this association, the empirical studies varied task complexity mainly through topping up functions \cite{Tang_Dai_2011} and increasing the load of mathematical calculations \cite{Ou_Pang_2012,Pu_Xu_2013}, while the simulation study \cite{Sheng_Hu_2014} characterized the complexity in computation algorithm as a random variable with Gamma distribution.

    	 \item	\textbf{Task Size:} As mentioned previously, a composite-object task can further be defined as a combination of the input/output data and computation workload \cite{Deng_Huang_2015}, and therefore the size of a task can partially be reflected by the data size \cite{Chen_Schneider_2012} or together with the computation complexity \cite{Sheng_Hu_2014}. To avoid duplication, we only focus on the amount of computation workload \cite{Saab_Chehab_2013} that has been widely depicted as the number of CPU cycles \cite{Balakrishnan_Tham_2013}, floating-point operations \cite{Mazza_Tarchi_2014,Mtibaa_Harras_2013}, and processing instructions \cite{Deng_Huang_2015,Fard_Prodan_2012,Kim_Buyya_2007,Kumar_Lu_2010,Xia_Liang_2014}. In fact, the CPU cycles of a computation task have been treated as a linear function of the data input the task \cite{Song_Cui_2014}, and the computation complexity can also be translated into particular number of instructions \cite{Kim_2012}.         
	\end{enumerate}

\subsubsection{Summary}
\label{subsubsec:workloadFactorDis}
\begin{figure}[!t]
\centering
\includegraphics[width=11.5cm]{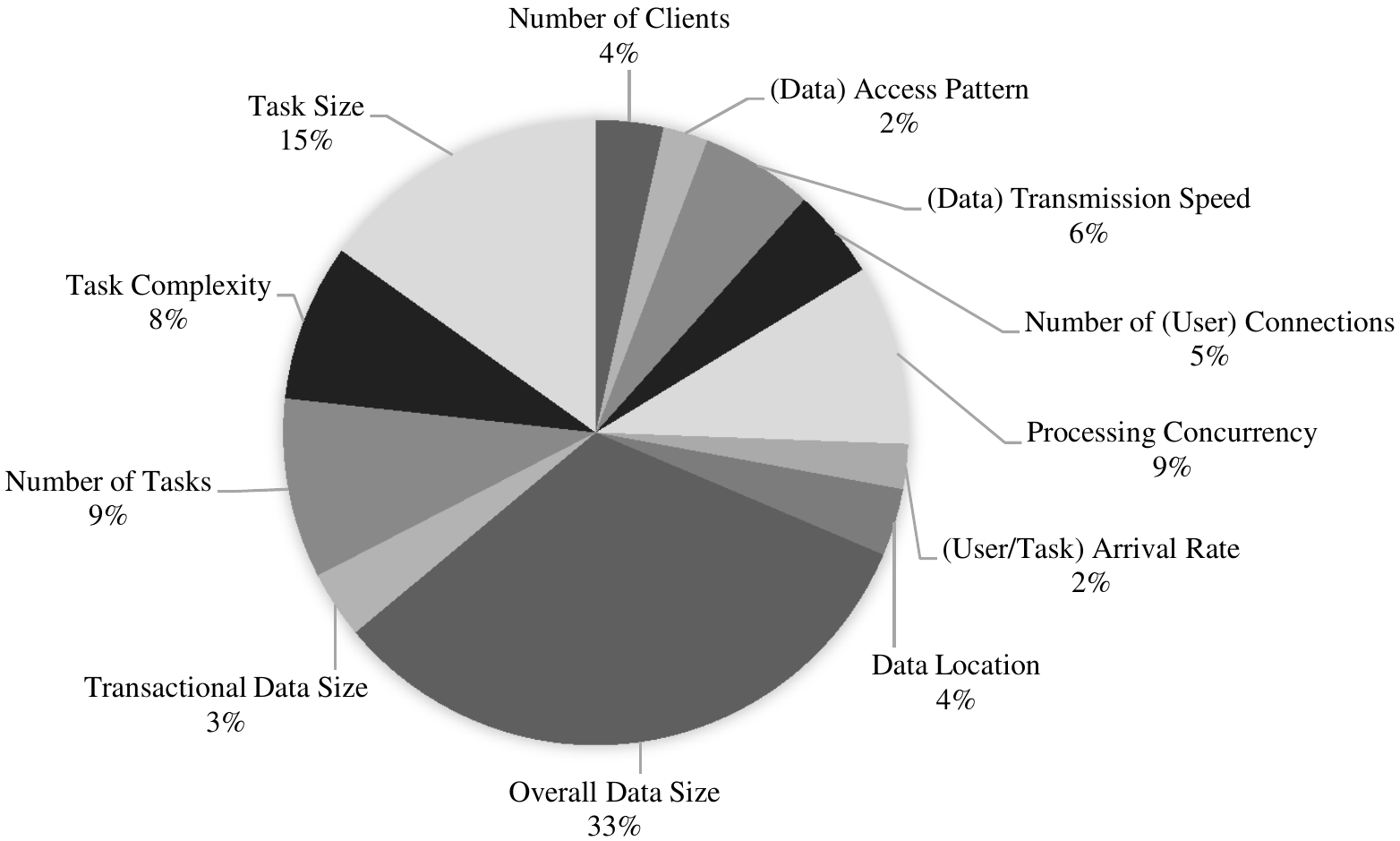}
\caption{Study popularity of workload factors that are influential on energy consumption of Cloud applications. (The total number of factor-studies is 86.)}
\label{PicApplicationFactors}
\end{figure}

As listed above, we have identified 12 workload factors in total. In a similar fashion to Section \ref{subsubsec:EnvironFactorDicussion}, we also use numerical factor-studies to reflect to what extent different environmental factors have concerned researchers, as illustrated in Fig.~\ref{PicApplicationFactors}. It is again notable that popular factors do not necessarily act as main contributors to energy consumption.

By isolating individual factors' impacts on energy consumption from each other, the sizes of data and task (the processing workload) seem to be the main energy-related factors in a Cloud application. In fact, there has been a wide consensus on these two factors among the literature and reality: Computational tasks rely on the major power consumer of computing resources \cite{Lefurgy_Wang_2008}, while the data lead to communication and storage costs. Such a factor concentration roughly matches the main environmental factors (cf.~Section \ref{subsubsec:EnvironFactorDicussion}) in terms of their potential interactions (i.e.~task processing and data communication).

Since Cloud application workload is usually reflected by a combination of factors, in practice, one factor's influence on energy consumption could be correlated with or even constrained by others. For example, task size and task complexity can sometimes interchangeably indicate each other (\cite{Sheng_Hu_2014} vs.~\cite{Kim_2012}); the number of tasks and data size are frequently used together to represent the overall workload size (e.g., \cite{Mazza_Tarchi_2014}); while the degree of parallelism in a Cloud application also depends on the resource allocations (e.g., \cite{Wirtz_Ge_2011}). We leave more discussions about combinational influences of factors to Section \ref{sec:futuredirections}.

\subsection{Energy Consumption Models of Cloud Applications (RQ5)}
 \label{subsec:RQ5}

\begin{table}\footnotesize %
\renewcommand{\arraystretch}{1.1}
\caption{Summary of Key Notations\label{tab:one}}{%
\begin{tabular}{|l| >{\raggedright\arraybackslash}p{11.5cm}|}
\hline
\textbf{Symbol}   & \textbf{Brief Explanation}\\
\hline
$a,k$     & Predefined constant coefficients.\\\hline
$A$     & The Cloud application.\\\hline
$C$ & Total CPU cycles as the computational workload involved in a particular task.\\
\hline
$D(\cdot)$     & Data size function either of Cloud application/tasks (e.g., $D(n_i)$), or of environmental resource/items (e.g., $D(r_i\to)$). In the former case, it represents the size of data involved in an application/task. In the latter case, it uses $\to$ or $\gets$ to indicate the size of data sent/received from/by a resource item.\\\hline
$\widehat D(\cdot)$     & The maximum content capacity (or size) of the memory or the hard disk.\\\hline
$e(\cdot)$  & Workload-oriented energy rate function of Cloud application/tasks (e.g., $e(n_i)$). Unlike $P(\cdot)$, it defines the energy expense during a unit of time when dealing with workloads. \\
\hline
$E(\cdot)$     & Energy consumption function of Cloud application/tasks (e.g., $E(A)$). It can use a superscript to specify the relevant resource (e.g., $E^\mathit{client}(\cdot)$), and a subscript to indicate the energy consumption component (e.g., $E_\mathit{active}(\cdot)$).\\\hline
$f,v$     & The operating frequency (i.e.~$f$) and supply voltage (i.e.~$v$).\\\hline
$M,N$     & The total number of environmental resource items (i.e.~$M$) and Cloud application tasks (i.e.~$N$).\\\hline
$n_i$     & The $i^\mathit{th}$ task of the Cloud application $A$. In particular, the subscript can be replaced with $\mathit{cpu}$, $\mathit{net}$, $\mathit{mem}$, or $\mathit{disk}$ to indicate a particular type of resource-intensive task.\\\hline
$P(\cdot)$     & Power consumption function of environmental resource/items (e.g., $P(r_i)$). It can further include $t$ or $\Phi(r_i)$ to indicate the power at a particular time point or data throughput (e.g., $P(r_i,t)$ or $P(r_i,\Phi(r_i))$). If needed, a subscript is used to specify the power consumption component (e.g., $P_\mathit{idle}(\cdot)$).\\\hline
$r_i$     & The $i^\mathit{th}$ resource item in a particular resource pool $R(\cdot)$. If needed, a particular resource and/or its component can further be specified in the subscript (e.g., $r_{\mathit{client},\mathit{cpu}}$).\\\hline
$R(\cdot)$     & Environmental resource function of Cloud application/tasks (e.g., $R(A)$).\\\hline
$S(\cdot)$     & Compute speed function of environmental resource/items (e.g., $S(r_i)$).\\\hline
$t$     & A particular time point.\\\hline
$T(\cdot)$     & Time span function either of Cloud application/tasks (e.g., $T(A)$), or of environmental resource/items (e.g., $T(r_i)$). In the latter case, it can use a subscript to imply the resource state (e.g., $T_\mathit{idle}(\cdot)$).\\
\hline
$U(t)$     & Resource utilization ratio at time point $t$.\\\hline
$W(\cdot)$     & Workload size function of Cloud application/tasks (e.g., $W(A)$). It can further include $t$ to indicate the workload at a particular time point (e.g., $W(n_i,t)$).\\\hline
$\alpha,\beta,\gamma,\lambda$     & Regression parameters that need to be determined by experimental measurements.\\\hline
$\delta$     & A particular fraction ratio.\\\hline
$\Theta$     & Data transmission channel quality with variable value $0<\Theta<1$.\\\hline
$\tau$     & Execution time of a particular task at the maximum processing capacity.\\\hline
$\Phi(\cdot)$     & Data throughput function of the channel (either network communication or data reading/writing) between two resource items (e.g., $D(r_i\to r_j)$). If needed, it is possible to emphasize one resource item only (e.g., $\Phi(r_i\gets)$), and also to ignore the data flow direction (e.g., $\Phi(r_i)$).\\
\hline
$\widehat\Phi(\cdot)$     & The maximum data throughput capacity (or bandwidth) between two resource items or of a single resource item.\\\hline
$\Omega$     & The set of power-consuming components contained in a particular resource item.\\\hline
\end{tabular}}
\end{table}%

Recall that it is extremely challenging to deal with energy-related issues of Cloud applications due to the inherent complexity in the applications themselves and the heterogeneity in their deployment environments \cite{Chen_Grundy_2015}. By abstracting energy consumption behaviors and details, the mathematical models have pervasively been employed to help understand and in turn investigate how the energy is consumed for running a Cloud application. To facilitate discussing, comparing, and reporting the identified energy consumption models, we unify the various notations from the relevant studies, as listed in Table \ref{tab:one}. Moreover, except for the models that hold implicit environmental views, we follow the previous environmental deconstruction to organize the identified models respectively representing overall energy consumption as well as computation, communication, and storage energy consumption of Cloud applications. In particular, we did not find memory-specific energy consumption models in the context of Cloud applications.

\subsubsection{Environment-implicit Energy Consumption Model}
As the name suggests, the environment-implicit energy consumption models are purely based on the analysis of Cloud applications, with little consideration of the deployment environment. Without loss of generality, we exploit the widely employed directed acyclic graph (DAG) as a generic model of Cloud application $A$ in our discussion, as shown in Equation (\ref{eqn:DAG}).

\begin{small}\begin{equation}
\label{eqn:DAG}
A: 
                \begin{cases}
                  \mathit{DAG} &= \{\mathit{Node},~ \mathit{Edge}\}\\
                  \mathit{Node} &= \{n_i ~|~ 1 \leq i \leq N\}\\
                  \mathit{Edge} &= \{(n_i, ~n_j)~ |~ n_i \in \mathit{Node}, ~n_j \in \mathit{Node}\}
                \end{cases}
\end{equation}\end{small}%
where the application's $\mathit{DAG}$ comprises $N$ nodes and at most $N\times N$ edges. By partitioning $A$ into functional pieces, each node $n_i$ indicates a workload task, while each edge $(n_i, ~n_j)$ represents the precedence constraint between two consecutive tasks. Unlike the application modeling in \cite{Liu_Cao_2014,Wu_Wolter_2015}, we treat data transmission as a workload task represented by a node instead of an edge.

By focusing only on the execution duration and the required energy unit of each workload task, the most straightforward energy consumption model of $A$ was given in \cite{Fard_Prodan_2012,Wajid_Marin_2013}: 
\begin{small}\begin{equation}
\label{eqn:MostAbstractModel}
E(A) = \sum_{i=1}^{N} e(n_i) \cdot T(n_i)
\end{equation}\end{small}%
where $E(\cdot)$ represents a generic energy consumption function, while $T(\cdot)$ is a generic makespan function. Note that $e(n_i)$ is the energy unit consumed by the task $n_i$ during a unit of time, which essentially is a workload-oriented notation \cite{Chen_Schneider_2012,Chen_Grundy_2013,Kim_2012} in contrast to the power consumption in environmental resources. In addition to the task energy per time unit, there are also other types of workload-oriented energy units, e.g., energy per user or energy per bit \cite{Vishwanath_Jalali_2015}.

When individual workload tasks have the same functionality, they can be grouped together to facilitate energy consumption modeling. For example, in the context of a MapReduce workflow, there are generally mapping, shuffling and reducing tasks. Correspondingly, the study \cite{Lin_Leu_2015} defined a \textbf{function-group-based energy consumption model} as:  
\begin{small}\begin{equation}
\label{eqn:MapReduce}
E(A) = 
                \begin{cases}
                  E(\mathit{map}) + E(\mathit{hold}) + E(\mathit{reduce}) &\text{~~if local data,}\\
                  E(\mathit{map}) + E(\mathit{replicate}) +E(\mathit{hold}) + E(\mathit{reduce}) &\text{~~if distributed data.} 
                \end{cases}
\end{equation}\end{small}%

Recall that there are mainly four types of infrastructural resources (cf.~Section \ref{subsec:RQ4-2}). Without necessarily knowing the environmental details, similarly, we can also group the tasks that are related to the same resource-intensive workload. As for the task interactions, their energy consumption comprises an integration of task computation and information communication between tasks \cite{Luo_Yang_2015}. Although few modeling studies were concerned with the four resource types simultaneously, we summarize such a \textbf{resource-group-based energy consumption model} inspired by the empirical investigation \cite{Zhang_Fu_2011}, as shown below.
\begin{small}\begin{equation}
\label{eqn:Resource}
E(A) = E(\mathit{communication}) + E(\mathit{computation}) +E(\mathit{memory}) + E(\mathit{storage})
\end{equation}\end{small}%

Since this model is inherently associated with Cloud applications' deployment environment, we further treat it as a bridge between the environment-implicit and the following environment-specific energy consumption models. 

\subsubsection{Environment-specific Overall Energy Consumption Model} 
\label{subsubsec:overallmodel}
When it comes to the environment of a Cloud application, we are only concerned with the IT equipment, while not including cooling and other facilities. From the viewpoint of resource partitioning, the deployment environment of Cloud applications has normally been modeled as a resource pool comprising a set of resource items:
\begin{small}\begin{equation}
\label{eqn:resourceItem}
R(A)= \{r_i ~|~ 1 \leq i \leq M\}
\end{equation}\end{small}%
where $r_i$ is the $i^{th}$ resource item within the pool $R(A)$ consisting of $M$ resource items. Since we focus on the environmental resources with respect to a single Cloud application, in this survey, we clarify that $R(A)$ is only composed of the resource items employed by the aforementioned Cloud application $A$. Moreover, the employed resource items might have different types \cite{Xiao_Hu_2013}, and the same type of resource items are not necessarily identical \cite{Thanavanich_Uthayopas_2013}. Then, the energy consumption of $A$ can be modeled based on the involved resources' power consumptions.
In fact, a key characteristic of environment-specific modeling is that it relies on the power consumption of environmental resources. For example, by denoting the power consumed in the resource item $r_i$ at time $t$ to be $P(r_i,t)$, the studies \cite{Ge_Feng_2010_TPDS,Wirtz_Ge_2011} modeled the energy expense of a parallel application $A$ running with $M$ resource items during time interval $(t_1,t_2)$:
\begin{small}\begin{equation}
\label{eqn:deltaA}
\Delta E(A)= \sum_{i=1}^{M}\int_{t_1}^{t_2}P(r_i,t)\cdot dt
\end{equation}\end{small}%

If we define every resource item to be a combination of various power-consuming components, $P(r_i,t)$ of resource $r_i$ can further be specified into $\sum_{j\in \Omega} P(r_{i,j},t)$, where $\Omega$ is the set of power-consuming components \cite{Xiao_Hu_2013}. By dividing $\Omega$ into the aforementioned four resource types (namely \textit{cpu}, \textit{net}, \textit{mem} and \textit{disk} for short), we are able to update Equation (\ref{eqn:deltaA}) and make it compatible with Equation (\ref{eqn:Resource}):
\begin{small}\begin{equation}
\label{eqn:deltaAupdated}
\begin{split}
\Delta E(A)=& \sum_{i=1}^{M}\int_{t_1}^{t_2}\sum_{j\in \Omega} P(r_{i,j},t)\cdot dt \\ =&\sum_{i=1}^{M}\int_{t_1}^{t_2}\left(P(r_{i,\mathit{cpu}},t)+P(r_{i,\mathit{net}},t)+ P(r_{i,\mathit{mem}},t)+ P(r_{i,\mathit{disk}},t)\right)\cdot dt
\end{split}
\end{equation}\end{small}%

If focusing on the CMOS circuits involved in the IT resources \cite{Huang_Su_2012}, since a CMOS circuit has two power consumption components (namely static power and dynamic power), a Cloud application's energy consumption can be distinguished between the static and dynamic parts \cite{Wu_Huang_2013}, as shown in Equation (\ref{eqn:twoPowerParts}).
\begin{small}\begin{equation}
\label{eqn:twoPowerParts}
E(A) = E_\mathit{static}(A) +E_\mathit{dynamic}(A) =\left(P_\mathit{static}(R(A))+P_\mathit{dynamic}(R(A))\right)\cdot T(A)
\end{equation}\end{small}%
where $P_\mathit{static}(R(A))$ and $P_\mathit{dynamic}(R(A))$ represent the average static and dynamic power consumed in the application environment $R(A)$ during the application runtime $T(A)$.

In theory, \textbf{Static Power} indicates the essential power for keeping IT resources in the power-on state (e.g., maintaining the basic circuits and system clock), which is independent of any workload \cite{Balakrishnan_Tham_2013} and cannot be avoided until the whole system is turned off \cite{Xiao_Hu_2013,Zhang_Li_2015}. As such, the static power consumption is normally modeled as a constant without scaling with other factors \cite{Abdelsalam_Maly_2009}. In practice, the reverse-bias leakage between diffused regions and the substrate will also result in a particular amount of static power consumption, while the leakage can be proportionally influenced by the temperature \cite{Chaparro_Gonzalez_2004}. Further considering the proportional impact of dynamic power on the temperature, some studies estimated the static power as a fraction of its dynamic counterpart, and the fraction is usually less than 30\% \cite{Kim_Buyya_2007,Wang_Laszewski_2010}. Thus, during the execution of an application, the static energy consumption can be expressed as:
\begin{small}\begin{equation}
\label{eqn:computeStatic}
E_\mathit{static}(A) = \delta\cdot E_\mathit{dynamic}(A),~~0\%<\delta<30\%
\end{equation}\end{small}%

\textbf{Dynamic Power} is the dynamic utilization of power in the environmental IT resources when dealing with workloads. Since the dynamic power dominates the whole power consumption in the popular CMOS technology \cite{Zhang_Li_2015}, most of the relevant studies only employed the dynamic power for modeling the energy consumption of Cloud applications (e.g., \cite{Balakrishnan_Tham_2013,Thanavanich_Uthayopas_2013}). 

Furthermore, from the perspective of a system rather than of a CMOS gate, we distinguish between the active and idle power consumption according to different load levels of a particular IT resource during the execution of a Cloud application \cite{Chun_Ihm_2011,Zheng_Huang_2014}. \textbf{Active Power} refers to the power for actively executing tasks on an IT resource (i.e.~$>0\%$ load), and \textbf{Idle Power} indicates the power consumption when the IT resource is ready to work while doing nothing (i.e.~$0\%$ load). Note that IT resources are not truly static at idle states, because there are still backend workloads.\footnote{\url{https://wiki.mcs.anl.gov/cqos/index.php?title=Power_Specifications_and_Model}} To be aligned with the definition of dynamic power (when dealing with workloads), we clarify that static power is excluded when discussing active power and idle power in this survey. In fact, the study \cite{Balakrishnan_Tham_2013} has combined idle power with static power (e.g., the power corresponding to the \textbf{sleep} resource state \cite{Hsu_Lin_2013,Namboodiri_Ghose_2013}) into the so-called \textbf{standby power}. Therefore, by focusing on the dynamic power, the dynamic energy expense for completing the Cloud application $A$ can be modeled as:
\begin{small}\begin{equation}
\label{eqn:dynamicPower}
\begin{split}
               E_\mathit{dynamic}(A) &= E_\mathit{idle}(A)+ E_\mathit{active}(A) \\
&= P_\mathit{idle}(R(A))\cdot T_\mathit{idle}(R(A)) +P_\mathit{active}(R(A))\cdot T_\mathit{active}(R(A))
\end{split}
\end{equation}\end{small}%
where $T_\mathit{idle}(R(A))$ and $T_\mathit{active}(R(A))$ respectively indicate the average idle and the average active time of the environmental IT resources $R(A)$. It is noteworthy that $T_\mathit{idle}(R(A))+T_\mathit{active}(R(A))\neq T(A)$. Since different resource items are possible to be alternatively idle during the continuous execution of the Cloud application $A$, it is improper to use fractions of $T(A)$ to calculate $A$'s idle and active energy consumption.

When it comes to $E_\mathit{active}(A)$, one of the active energy components reflects the energy used for driving the data flow of the Cloud application $A$. The data flow might comprise various interactive execution elements (cf.~Fig.~\ref{PicActivities}) with respect not only to network equipment (e.g., \cite{Balakrishnan_Tham_2013}) but also to other types of resources (e.g., \cite{Xiao_Hu_2013}). From the perspective of a single resource item $r_i$, the corresponding data flow can be distinguished as either data input or data output. By emphasizing the input/output channel between two consecutive resource items, the energy consumption of $A$'s data flow has been modeled as follows.
\begin{small}\begin{equation}
\label{eqn:twoResourceCommunication}
E_\mathit{dataflow}(A) =\sum_{r_i, r_j \in R(A)} \left(P_\mathit{out}(r_i) + P_\mathit{in}(r_j)\right)\cdot \frac{D(r_i{\to}r_j)}{\Phi(r_i{\to}r_j)}, ~~i\neq j
\end{equation}\end{small}%
where $P_\mathit{out}(r_i)$ (resp.~$P_\mathit{in}(r_j)$) is the power of resource $r_i$ (resp.~$r_j$) when outputting/inputting the data $D(r_i{\to}r_j)$, and $\Phi(r_i{\to}r_j)$ refers to the data throughput between those two different resource items $r_i$ and $r_j$. 

Instead of emphasizing the input/output channel, Equation (\ref{eqn:twoResourceCommunication}) has been rewritten in \cite{Song_Cui_2014} by focusing on the input/output activities of individual resource items:
\begin{small}\begin{equation}
\label{eqn:oneResourceCommunication}
E_\mathit{dataflow}(A) =\sum_{r_i \in R(A)} \left(P_\mathit{out}(r_i)\cdot \frac{D(r_i{\to })}{\Phi(r_i{\to})} + P_\mathit{in}(r_i)\cdot \frac{D(r_i{\gets})}{\Phi(r_i{\gets})}\right)
\end{equation}\end{small}%
where $D(r_i{\to})$/$D(r_i{\gets})$ represents the size of output/input data of the resource item $r_i$, and $\Phi(r_i{\to})$/$\Phi(r_i{\gets})$ indicates the data throughput when $r_i$ is outputting/inputting data. 

\subsubsection{Environment-specific Computation Energy Consumption Model}
\label{subsubsec:computeEnvironmentModel}
Following the convention of Equation (\ref{eqn:MostAbstractModel}) and (\ref{eqn:resourceItem}), here we consider a computation-intensive task $n_\mathit{cpu}$ running on the compute resource $r_\mathit{cpu}$. As mentioned above, the dynamic power dominates the power consumption of CPU's CMOS circuits, and the dynamic CPU power generally depends on the supply voltage and operating frequency via relation $P_\mathit{dynamic}(r_\mathit{cpu})=k\cdot v^2\cdot f$ \cite{Wu_Wu_2015}. The operating frequency-based model specified in Equation (\ref{eqn:DVFS}) has widely been used for applications' energy consumption in both client devices and Cloud servers.  
\begin{small}\begin{equation}
\label{eqn:DVFS}
E_\mathit{dynamic}(n_\mathit{cpu}) = k\cdot v^2\cdot f\cdot T(n_\mathit{cpu}) =k\cdot a^2\cdot f^3\cdot T(n_\mathit{cpu}) 
\end{equation}\end{small}%
where the energy coefficient $k$ depends on the CPU's chip architecture; the linearly proportional relationship between the operating clock frequency $f$ and the supply voltage $v$ is modeled as $v=af$; while $a$ is a constant coefficient. 

It is evident that the consumed energy of the task $n_\mathit{cpu}$ is directly proportional to its makespan, i.e. $E(n_\mathit{cpu}) \propto T(n_\mathit{cpu})$ \cite{Fekete_Csorba_2012}. However, a task's makespan varies in practice due to the dynamic changes in CPU capacity caused by possible voltage scaling at runtime. If using $\tau$ to denote the time for executing the task $n_\mathit{cpu}$ at the maximum processing capacity, then the practical execution time $T(n_\mathit{cpu})$ would be $\tau\cdot \frac{v_\mathit{max}}{v}$ \cite{Wu_Wu_2015} or $\tau\cdot \frac{f_\mathit{max}}{f}$ \cite{Kim_Beloglazov_2009}. In particular, the levels of voltage $v$ and frequency $f$ are within range $[v_\mathit{min}, v_\mathit{max}]$ and $[f_\mathit{min}, f_\mathit{max}]$ respectively. Accordingly, the previous frequency-based energy consumption model has been updated by \cite{Kim_Beloglazov_2009} into:
\begin{small}\begin{equation}
\label{eqn:DVFS-updated}
E_\mathit{dynamic}(n_\mathit{cpu}) = \int_{0}^{\tau\cdot \frac{f_\mathit{max}}{f}} k\cdot a^2\cdot f^3\cdot dt =k\cdot a^2\cdot f_\mathit{max}\cdot f^2\cdot \tau  
\end{equation}\end{small}%

Recall that the computation workload induced by a task can be measured by CPU cycles (cf.~Task Size in Section \ref{subsec:RQ4-1}). Suppose the task $n_\mathit{cpu}$ comprises $C$ cycles in total. Its makespan can directly be calculated as $C/f$ at frequency $f$. Then, as proposed in \cite{Kim_Buyya_2007,Ma_Lin_2015,Song_Cui_2014}, the energy consumption of such a task can be modeled as:
\begin{small}\begin{equation}
\label{eqn:cycle}
E_\mathit{dynamic}(n_\mathit{cpu}) =  k\cdot a^2\cdot f^3\cdot \frac{C}{f} =  k\cdot a^2\cdot f^2\cdot C 
\end{equation}\end{small}%

In the extreme case, the operating frequency is assumed changeable after every single CPU cycle \cite{Sheng_Hu_2014,Sheng_Mahapatra_2015}. Given the single cycle time $1/f_c$ at frequency $f_c$, one CPU cycle's energy consumption can be represented as $E(\mathit{cycle})=k\cdot a^2\cdot f_c^3\cdot \frac{1}{f_c} =k\cdot a^2\cdot f_c^2$, and thus the task's energy consumption can be expressed as:
\begin{small}\begin{equation}
\label{eqn:cycleUpdated}
E_\mathit{dynamic}(n_\mathit{cpu}) = \sum_{c=1}^{C} k\cdot a^2\cdot f_c^2 
\end{equation}\end{small}%

Considering that $f_c \in [f_\mathit{min}, f_\mathit{max}]$ and there are only limited frequency levels within $[f_\mathit{min}, f_\mathit{max}]$, we can categorize the CPU cycles into different frequency level groups. By using $\delta _f$ to denote the execution fraction of the task $n_\mathit{cpu}$ at the frequency $f$ \cite{Balakrishnan_Tham_2013}, the energy consumption model can be rewritten with regards to either the CPU cycles fractions (i.e.~$C\cdot \delta _f$) or the execution time fractions (i.e.~$T(n_\mathit{cpu})\cdot \delta _f$), as shown below. 
\begin{small}\begin{equation}
\label{eqn:frequencyLevel}
E_\mathit{dynamic}(n_\mathit{cpu}) = \sum_{f \in [f_\mathit{min}, f_\mathit{max}]} k\cdot a^2\cdot f^2\cdot C\cdot \delta _f =  \sum_{f \in [f_\mathit{min}, f_\mathit{max}]} k\cdot a^2\cdot f^3\cdot T(n_\mathit{cpu})\cdot \delta _f
\end{equation}\end{small}%

As explained in Equation (\ref{eqn:dynamicPower}), the idle state of compute resources caused by a task is generally unavoidable due to workload offloading or imbalanced parallel execution. In particular, a compute resource is considered to be idle when its operating frequency (or supply voltage) reaches the lowest level $f_\mathit{min}$ (or $v_\mathit{min}$) \cite{Huang_Su_2012}.  Accordingly, by focusing on the dynamic power, the dynamic energy expense for running the task $n_\mathit{cpu}$ on the resource $r_\mathit{cpu}$ can be separated and modeled as follows.
\begin{small}\begin{equation}
\label{eqn:cpuDynamic}
                \begin{cases}
                E_\mathit{idle}(n_\mathit{cpu}) &= k\cdot a^2\cdot f_\mathit{min}^3\cdot T_\mathit{idle}(r_\mathit{cpu})\\
                  E_\mathit{active}(n_\mathit{cpu}) &= \sum_{f \in (f_\mathit{min}, f_\mathit{max}]} k\cdot a^2\cdot f^3\cdot T_\mathit{active}(r_\mathit{cpu})\cdot \delta _f  
              
                \end{cases}
\end{equation}\end{small}%

Similar to Equation (\ref{eqn:deltaA}) and (\ref{eqn:deltaAupdated}), it is also common to model Cloud application energy consumption without specifying the power details such as operating frequency. For example, by assuming the resource power $P(r_\mathit{cpu})$ and the compute speed $S(r_\mathit{cpu})$ to be constant when running the task $n_\mathit{cpu}$, the consumed energy was calculated in \cite{Xia_Liang_2014} through:
\begin{small}\begin{equation}
\label{eqn:computePower}
E_\mathit{active}(n_\mathit{cpu}) = P_\mathit{active}(r_\mathit{cpu}) \cdot \frac{W(n_\mathit{cpu})}{S(r_\mathit{cpu})}  
\end{equation}\end{small}%
where $W(\cdot)$ is a generic workload function, and then $W(n_\mathit{cpu})$ refers to the workload of the task $n_\mathit{cpu}$. It is clear that the idle state of compute resource has been excluded in this case. Therefore, we particularly label Equation (\ref{eqn:computePower}) as an active energy consumption model.

Instead of a constant value, the power consumed in a compute resource has been identified to be an exponential function of the resource utilization \cite{Tang_Dai_2011}. By using $P_\mathit{idle}(r_\mathit{cpu})$ and $P_\mathit{full}(r_\mathit{cpu})$ to respectively represent the compute resource's empty and full load powers, the energy consumption for running the task $n_\mathit{cpu}$ on the compute resource can be modeled as:
\begin{small}\begin{equation}
\label{eqn:utilization}
E_\mathit{dynamic}(n_\mathit{cpu}) = \int_{0}^{T(n_\mathit{cpu})} \left( P_\mathit{idle}(r_\mathit{cpu}) + \left(P_\mathit{full}(r_\mathit{cpu})-P_\mathit{idle}(r_\mathit{cpu})\right)\cdot \alpha\cdot U(t)^\beta\right)\cdot dt   
\end{equation}\end{small}%
where $\alpha$ and $\beta$ are resource-specific parameters that need to be determined through empirical measurements. The context-dependent notation $U(t)$ denotes the utilization of compute resource at time $t$. In the straightforward case, $U(t)$ directly equals to the CPU load fraction \cite{Namboodiri_Ghose_2013}. As for a multi-CPU server, $U(t)$ was estimated as the number of active CPU cores among all the available ones \cite{Enokido_Takizawa_2015}. Considering that the compute resource utilization would also be proportional to the workload being dealt with, the study \cite{Tang_Dai_2011} further modeled $U(t)=\gamma \cdot W(n_\mathit{cpu},t) + \lambda$, where $\gamma$ and $\lambda$ are both resource-specific parameters, and the workload $W(n_\mathit{cpu},t)$ was measured by the number of user connections at time $t$.

\subsubsection{Communication Energy Consumption Model}
\label{subsubsec:communicationModel}
Similarly, we define a communication-intensive task $n_\mathit{net}$ of $A$ to facilitate our discussion. As explained in Section \ref{subsubsec:overallmodel}, the task $n_\mathit{net}$ can be thought of as a data flow across the involved resource pool $R(n_\mathit{net})$, and then $E(n_\mathit{net})$ can directly be derived from Equation (\ref{eqn:twoResourceCommunication}) and (\ref{eqn:oneResourceCommunication}) \cite{Balakrishnan_Tham_2013,Song_Cui_2014}. 

Given the generic architecture for physical environment of Cloud applications (cf.~Section \ref{subsec:environment}), the resource items can be grouped into Client, Internet, Cloudlet, and Cloud resources. Accordingly, the communication energy consumption of a Cloud application can roughly be divided into four parts \cite{Altamimi_Naik_2011}, as modeled as follows. 
\begin{small}\begin{equation}
\label{eqn:communication}
E(n_\mathit{net}) =  E^\mathit{client}(n_\mathit{net}) + E^\mathit{internet}(n_\mathit{net})
  + E^\mathit{cloudlet}(n_\mathit{net})+ E^\mathit{cloud}(n_\mathit{net})
\end{equation}\end{small}%

It is noteworthy that Equation (\ref{eqn:twoResourceCommunication}) and (\ref{eqn:oneResourceCommunication}) are still valid and can be reused for each of the four energy parts by adapting the resource pool.

As the most controllable part, the client side attracts most of the research efforts on modeling communication energy consumption. By treating a client device as a single resource item, a straightforward approach is to follow Equation (\ref{eqn:oneResourceCommunication}) to estimate the communication energy consumed in client devices, as follows. 
\begin{small}\begin{equation}
\label{eqn:oneResourceCommunicationDeduce}
E^\mathit{client}(n_\mathit{net}) =P_\mathit{send}(r_\mathit{client,net})\cdot \frac{D(r_\mathit{client}{\to })}{\Phi(r_\mathit{client}{\to})} + P_\mathit{receive}(r_\mathit{client,net})\cdot \frac{D(r_\mathit{client}{\gets})}{\Phi(r_\mathit{client}{\gets})}
\end{equation}\end{small}%

Without distinguishing the power \cite{Kumar_Lu_2010} and data \cite{Xia_Liang_2014} between sending and receiving, Equation (\ref{eqn:oneResourceCommunicationDeduce}) can be simplified to:
\begin{small}\begin{equation}
\label{eqn:deviceCommunication}
\begin{split}
E^\mathit{client}(n_\mathit{net}) &= P(r_\mathit{client,net})\cdot\frac{D(n_\mathit{net})}{\Phi(r_\mathit{client})} \text{~~~~or}\\
E^\mathit{client}(n_\mathit{net}) &= P(r_\mathit{client,net})\cdot\frac{2\cdot D(n_\mathit{net})}{\Phi(r_\mathit{client})}
\end{split}
\end{equation}\end{small}%
where $P(r_\mathit{client,net})$ and $\Phi(r_\mathit{client})$ are respectively the transmission power and data throughput of the client device $r_\mathit{client}$. Note that we use $r_\mathit{client,net}$ to emphasize the power consumed in the network component of the resource $r_\mathit{client}$; and the notation $\Phi(r_\mathit{client})$ completely ignores the data transmission directions. As such, the first expression in Euqation (\ref{eqn:deviceCommunication}) views $D(n_\mathit{net})$ as the overall roundtrip data in the task $n_\mathit{net}$, while in the second expression $D(n_\mathit{net})$ is doubled to imply the data transmission along both directions. 

Considering the influence of uncertain channel quality (e.g., transmission errors), the factor Network Condition (cf.~Section \ref{subsec:RQ4-2}) was introduced to the previous cases \cite{Ravi_Peddoju_2015}:
\begin{small}\begin{equation}
\label{eqn:deviceConditionCommunication}
E^\mathit{client}(n_\mathit{net}) = P(r_\mathit{client,net})\cdot\left(\frac{ D(r_\mathit{client}{\to})}{\Phi(r_\mathit{client}{\to})}+\beta_1 + \frac{ D(r_\mathit{client}{\gets})}{\Phi(r_\mathit{client}{\gets})}+\beta_2\right)
\end{equation}\end{small}%
where $\beta_1$ and $\beta_2$ are the channel condition parameters for sending and receiving data respectively, and their values are required to be tested by the client device $r_\mathit{client}$ itself \cite{Ravi_Peddoju_2015}. We note that, in this model, the data sending and receiving power of $r_\mathit{client}$ are assumed to be identical. By using regression analysis and Wolfram Mathematica, the study \cite{Akram_ElNahas_2015} even ignored the data transmission power, and proposed the following energy consumption model:
\begin{small}\begin{equation}
\label{eqn:deviceNopowerCommunication}
E^\mathit{client}(n_\mathit{net}) = \frac{\alpha \cdot D(n_\mathit{net}) - \beta}{\Phi(r_\mathit{client})}
\end{equation}\end{small}%
where $\alpha$ and $\beta$ are constant parameters that need to be determined through experimental measurements. Resorting to the Shannon Formula, $\Phi(r_\mathit{client})$ was further modeled as $\Phi(r_\mathit{client})=\frac{\widehat\Phi(r_\mathit{ap})}{\mathit{number~of~clients}}\cdot\log_2 \left(1+\frac{\mathit{SNR}}{\mathit{Distance}(r_\mathit{client},r_\mathit{ap})^2}\right)$, with regarding to the signal to noise ratio $\mathit{SNR}$, the bandwidth $\widehat\Phi(r_\mathit{ap})$ and resource competition of the access point $r_\mathit{ap}$, and the distance between $r_\mathit{ap}$ and $r_\mathit{client}$ \cite{Mazza_Tarchi_2014}.

By replacing transmission throughput with channel quality, the studies \cite{Sheng_Hu_2014,Sheng_Mahapatra_2015} proposed the following convex monomial function to describe the energy used to transmit $D(n_\mathit{net})$ bits of data:
\begin{small}\begin{equation}
\label{eqn:channelStateCommunication}
E^\mathit{client}(n_\mathit{net}) = \gamma \cdot \frac{D(n_\mathit{net})^o}{\Theta}
\end{equation}\end{small}%
where $\gamma$ denotes the energy coefficient in the order of less than $10^{-2}$, $\Theta$ represents the channel state with variable value $0<\Theta<1$ at different time slots, and $o$ refers to the order of monomial that depends on the transmission scheduling policy. For instance, the one-shot policy $o=1$ is used to indicate that the channel state has the biggest influence on the data transmission, and the transmission is finished in one time slot only.

Without conflicting with such a one-shot policy, a further simplified model proposed a directly proportional relation between the energy consumption of a communication task and its data size, i.e.~$E^\mathit{client}(n_\mathit{net})\propto D(n_\mathit{net})$ \cite{Fekete_Csorba_2012,Segata_Bloessl_2014}, as shown below:
\begin{small}\begin{equation}
\label{eqn:dataCommunication}
E^\mathit{client}(n_\mathit{net}) = \lambda \cdot D(n_\mathit{net})
\end{equation}\end{small}%
where $\lambda$ is a linear or quantile regression parameter that can be related to the employed access point technology \cite{Segata_Bloessl_2014}.

By analogy with CMOS concern, the network power of client devices, $E^\mathit{client}(n_\mathit{net})$, can also be separated into static part and dynamic part \cite{Wu_Huang_2013}, where the dynamic part covers the idle and active states \cite{Namboodiri_Ghose_2013}. In particular, the active energy for wireless communication between the mobile device's RF module and different access points (cellular vs.~WiFi) was emphasized by \cite{Ma_Lin_2015,Xiang_Lin_2015}, as modeled below. To save space, here we replace the task $n_\mathit{net}$ with a dot.
\begin{small}\begin{equation}
\label{eqn:wirelessConnection}
E_\mathit{active}^\mathit{client}(\cdot) =             
	\begin{cases}
               E_\mathit{ramp}^\mathit{RF}(\cdot) + E_\mathit{transmit}^\mathit{RF}(\cdot) + E_\mathit{hold}^\mathit{RF}(\cdot) +E_\mathit{tail}^\mathit{RF}(\cdot) &\text{ ~~if cellular,} \\
                E_\mathit{scan}^\mathit{RF}(\cdot) + E_\mathit{transmit}^\mathit{RF}(\cdot) + E_\mathit{hold}^\mathit{RF}(\cdot) &\text{ ~~if WiFi.}
                \end{cases}
\end{equation}\end{small}%
where $E_\mathit{ramp}^\mathit{RF}(\cdot)$ refers to the extra energy for switching the RF circuitries from low- to high-power states before the initiation of cellular data transmission; $E_\mathit{tail}^\mathit{RF}(\cdot)$ indicates the tail energy of high-power duration after the cellular data transmission ends; $E_\mathit{scan}^\mathit{RF}(\cdot)$ represents the energy for scanning and associating to an available WiFi access point; $E_\mathit{transmit}^\mathit{RF}(\cdot)$ includes both the uplink and the downlink data transmission energy \cite{Deng_Huang_2015} that can be calculated through Equation (\ref{eqn:oneResourceCommunicationDeduce}), and the power value and data throughput need to be adapted to the chosen access point technology; while $E_\mathit{hold}^\mathit{RF}(\cdot)$ is the energy for keeping the access point interface active during the data transmissions.

Besides the client-side wireless network, the Internet was studied as another communication part for mobile Cloud applications in \cite{Luo_Yang_2015}. The communication energy consumed in the Internet was identified to be relative to the data size, the traffic load ratio and the transmission delay. However, the negative correlation between the transmission delay and the corresponding energy consumption conflicts with the other relevant studies and seems to be incorrect, thus our survey does not include the model proposed in \cite{Luo_Yang_2015}. 

By focusing on the routers only in the network path of a Cloud application, the study \cite{Izadpanah_Pawlikowski_2013} simplified the Internet architecture, and used the number of routers and their power profiles to model the data transmission energy:
\begin{small}\begin{equation}
\label{eqn:router}
E^\mathit{internet}(n_\mathit{net}) =\sum_{r_\mathit{router} \in R(n_\mathit{net})} P(r_\mathit{router},\Phi(r_\mathit{router}))\cdot \frac{D(n_\mathit{net})}{\Phi(r_\mathit{router})} 
\end{equation}\end{small}%
where $P(r_\mathit{router},\Phi(r_\mathit{router}))$ represents the power of the router $r_\mathit{router}$ at the data throughput $\Phi(r_\mathit{router})$, which implies that the router's power varies depending on its traffic load.

In practice, given different network segments of the Internet, the routers can be specified and classified according to their functions and locations, such as broadband gateway routers and edge/core routers. Moreover, the network path of a Cloud application also includes other types of network facilities like Ethernet switches and WDM transport equipment \cite{Vishwanath_Jalali_2015}. In detail, the user traffic over the Internet has been assumed to generally require three hops (over two switches, one broadband gateway router, and one edge router) before reaching the core network, and eight hops (over eight WDM links across nine core routers) within the core network \cite{Baliga_Ayre_2011}:
\begin{small}\begin{equation}
\label{eqn:Internet}
\begin{split}
E^\mathit{internet}(n_\mathit{net}) = 4\cdot\left(\frac{2\cdot P(r_\mathit{switch})}{\widehat\Phi(r_\mathit{switch})} + \frac{P(r_\mathit{broad})}{\widehat\Phi(r_\mathit{broad})} + \frac{P(r_\mathit{edge})}{\widehat\Phi(r_\mathit{edge})}+ \right.\\
\left.\frac{2\cdot 9\cdot P(r_\mathit{core})}{\widehat\Phi(r_\mathit{core})}+ \frac{8\cdot P(r_\mathit{wdm})}{2\cdot \widehat\Phi(r_\mathit{wdm})}\right)\cdot D(n_\mathit{net})
\end{split}
\end{equation}\end{small}%
where $P(r_\mathit{switch})$, $P(r_\mathit{broad})$, $P(r_\mathit{edge})$, $P(r_\mathit{core})$, and $P(r_\mathit{wdm})$ refers to the powers consumed in the Ethernet switch, broadband gateway router, edge router, core router, and WDM link respectively; and $\widehat\Phi(\cdot)$ represents the maximum capacity (or bandwidth) of the corresponding network equipment. The number of core routers are doubled to reflect the hardware redundancy of the core network, while the number of WDM links are halved to reflect the core hops between co-located equipment. The overall factor of four further covers extra power consumption under the redundancy policy (factor of 2) and high power expenditure at low network utilization (factor of 2). Note that we removed the factor of 1.5 for cooling and other overheads from the original study.

Similarly, by assuming two hops (over one switch, one edge router, and one gateway router) for accessing a server within a data center \cite{Altamimi_Naik_2011,Baliga_Ayre_2011}, the energy consumption of user traffic with respect to both the Cloudlet and the Cloud can be modeled as:
\begin{small}\begin{equation}
\label{eqn:Cloud}
E^\mathit{cloud}(n_\mathit{net}) = E^\mathit{cloudlet}(n_\mathit{net}) = 4\cdot\left(\frac{P(r_\mathit{switch})}{\widehat\Phi(r_\mathit{switch})} +  \frac{P(r_\mathit{edge})}{\widehat\Phi(r_\mathit{edge})}+ \frac{P(r_\mathit{gateway})}{\widehat\Phi(r_\mathit{gateway})}\right)\cdot D(n_\mathit{net})
\end{equation}\end{small}%
where $P(r_\mathit{gateway})$ and $\widehat\Phi(r_\mathit{gateway})$ respectively indicate the power and the maximum capacity of the gateway router.

\subsubsection{Storage Energy Consumption Model} 

Given a storage-intensive task $n_\mathit{disk}$, in addition to the data input/output analysis \cite{Xiao_Hu_2013} in alignment with Equation (\ref{eqn:twoResourceCommunication}), the major concern is about accessing data stored in hard disk arrays through content servers \cite{Baliga_Ayre_2011}. Naturally, the energy consumption of $n_\mathit{disk}$ can be split into two parts occurred in the disk arrays (i.e.~$E^\mathit{array}(n_\mathit{disk})$) and content servers (i.e.~$E^\mathit{server}(n_\mathit{disk})$) respectively:
\begin{small}\begin{equation}
\label{eqn:storage}
E(n_\mathit{disk}) = E^\mathit{array}(n_\mathit{disk}) + E^\mathit{server}(n_\mathit{disk})
\end{equation}\end{small}%

Suppose the data $D(n_\mathit{disk})$ involved in, or to be accessed by, the task $n_\mathit{disk}$ are pre-stored in the disk array $r_\mathit{array}$ (for the case of writing, we assume that the same size of storage area has been pre-booked in the disk array). Then, the energy for storing the data during the lifecycle $T(n_\mathit{disk})$ of the task can be calculated through:
\begin{small}\begin{equation}
\label{eqn:diskArrays}
E^\mathit{array}(n_\mathit{disk})=2\cdot D(n_\mathit{disk})\cdot \frac{P(r_\mathit{array})}{\widehat D(r_\mathit{array})}  \cdot T(n_\mathit{disk})
\end{equation}\end{small}%
where $P(r_\mathit{array})$ indicates the power of the hard disk array; $\widehat D(r_\mathit{array})$ stands for the disk content capacity; and the initial factor of 2 accounts for the redundancy policy in storage. As before, we removed the factor of 1.5 that reflects cooling and extra overheads for the power of the hard disk array.

For the purpose of conciseness, we define each task $n_\mathit{disk}$ to include only a one-shot access to the data $D(n_\mathit{disk})$, and multiple data accesses can be viewed as multiple tasks. Then, the data accessing energy consumed in a content server $r_\mathit{server}$ has been modeled by focusing either on the accessing time \cite{Izadpanah_Pawlikowski_2013} or on the data size \cite{Baliga_Ayre_2011}:
\begin{small}\begin{equation}
\label{eqn:contentServerStorage}
E^\mathit{server}(n_\mathit{disk})=P(r_\mathit{server,disk}) \cdot T(n_\mathit{disk})
= D(n_\mathit{disk})\cdot \frac{ P(r_\mathit{server,disk})}{\widehat\Phi(r_\mathit{server})} 
\end{equation}\end{small}%
where $P(r_\mathit{server,disk})$ refers to the power consumed in the storage component of  $r_\mathit{server}$, and $\widehat\Phi(r_\mathit{server})$ represents the maximum data throughput over $r_\mathit{server}$. In particular, the factor of extra power requirement for other overheads can also be added to Equation (\ref{eqn:contentServerStorage}) \cite{Altamimi_Naik_2011}.

If allowing multiple clients to access data simultaneously within the same task $n_\mathit{disk}$, the energy consumption located at $r_\mathit{server}$ between time $t_1$ and $t_2$ was given in \cite{Enokido_Suzuki_2010_AINA,Enokido_Suzuki_2010_CISIS} without emphasizing the storage component:
\begin{small}\begin{equation}
\label{eqn:contentServer}
\Delta E^\mathit{server}(n_\mathit{disk}) = \int_{t_1}^{t_2} \alpha\cdot \left(P_\mathit{idle}(r_\mathit{server}) + \beta_t \cdot\Phi(r_\mathit{server},t)\right)\cdot dt
\end{equation}\end{small}%
where $\alpha$ depends on the content server type, $\beta_t\geq 1$ is proportional to the number of clients at time $t$, and $\Phi(r_\mathit{server},t)$ refers to the data throughput over $r_\mathit{server}$ at time $t$.


\subsubsection{Summary}
Given the identified 30+ models, it is evident that there is no one-size-fits-all approach to modeling energy consumption of Cloud applications. Various energy consumption models are applied to different situations when emphasizing and combining different factors. By deconstructing and analyzing the existing models, however, we see a regular pattern in the modeling efforts, i.e.~on the power characteristics of the resources together with the way resources are utilized by application workloads. This regular pattern confirms the statement that a Cloud application's energy consumption involves a mutual effect between its workload and environmental factors. 

Furthermore, by distinguishing between different power consumption components, we see three viewpoints about the energy consumption of Cloud applications, and we name them as Effective, Active, and Incremental energy consumptions.

\textbf{Effective Energy Consumption}, i.e. $E(A)=E_\mathit{active}(R(A))+E_\mathit{idle}(R(A))$, includes both the active and the idle power consumed in the environmental resources of a Cloud application. In particular, the idle power consumption is included for two reasons. Firstly, the idle Cloud resources would have to keep a standby state and wait for new jobs, so that they can be rented again at any time \cite{Zheng_Huang_2014}. Secondly, the idle power consumption will still be meaningful and effective if it is used for maintaining the application accessibility and/or the data availability \cite{Barroso_Holzle_2007}.

\textbf{Active Energy Consumption}, i.e. $E(A)=E_\mathit{active}(R(A))$, includes only the active power consumed in the environmental resources of a Cloud application. Although the idle power consumption should not be excluded in practice as mentioned above, focusing on the active power consumption would be useful for investigating the energy consumption incurred by dynamic application activities.

\textbf{Incremental Energy Consumption}, i.e. $E(A)=E_\mathit{active}(R(A))+E_\mathit{idle}(R(A))-P_\mathit{idle}(R(A))\cdot T(A)$, is related to the increased power arising from the idle power consumed in the environmental resources of a Cloud application. In other words, the arising power consumption is the top-up part within active power consumption based on its idle counterpart. Since various IT equipment has widely different dynamic power ranges (e.g., network devices operating at the utilization less than 50\% may still incur nearly the maximum power consumption) \cite{Baliga_Ayre_2011,Barroso_Holzle_2007}, emphasizing the incremental energy consumption can reduce possible investigation bias by excluding the background noises \cite{Jalali_Gray_2014,Wirtz_Ge_2011}.


\section{Trade-off Debates}
\label{sec:futuredirections}
As mentioned in Section \ref{subsec:RQ4-1} and \ref{subsec:RQ4-2}, we try to isolate the influences of individual factors on the energy consumption, to avoid the combinatorial explosion of the factorial discussions. However, it is noteworthy that the energy expense of a Cloud application is inevitably affected as a result of combining multiple factors, as demonstrated in the mathematical models (cf.~Section \ref{subsec:RQ5}). Although studying various combinational factors' effects on the energy consumption is out of the scope of this survey, we particularly highlight a set of trade-off debates that would be worth further investigations, and we believe that model-based simulations would be the key to investigating those concerns raised by these debates.

\begin{itemize}

\item \textbf{Resource Allocation Level:} To improve the energy efficiency for a Cloud application, there is evidence advocating both less than and more than enough resource allocations. By provisioning ``under-the-just-enough" servers, the authors \cite{Guyon_Orgerie_2015} showed that a data-intensive Cloud application can save up to 24\% in energy consumption with a loss of around 6\% only in execution time. However, in general cases, Cloud applications are supposed to achieve greater energy efficiency by utilizing more processors, in order to finish more quickly and free the processors sooner. In other words, the energy saving for a Cloud application can be realized by returning its environmental resources to the idle state earlier \cite{Bonner_Namin_2012}.   

\item \textbf{Degree of Application Parallelism:} This debate can be viewed as a counterpart of the above one from the application's perspective. By tailoring the resource allocations to the degree of parallelism \cite{Wirtz_Ge_2011}, the overall energy consumption can decrease significantly with improved processing concurrency in a Cloud application \cite{Zhang_Fu_2011}. This is because the increased parallelism would have more chances to reduce the processing time and overwhelm the influence of the resource increase \cite{Wang_Laszewski_2010}. Nevertheless, considering the theoretical limit of energy saving of parallel executions \cite{Bonner_Namin_2012}, it is impossible to infinitely enhance the energy efficiency of a Cloud application by increasing its parallelism degree, not to mention that the increased overhead of process scheduling would meanwhile cause more energy consumption \cite{Chen_Schneider_2012,Chen_Grundy_2013}.

\item \textbf{Downscaling CPU Frequency:} In addition to the conflicting opinions on the effectiveness of adjusting CPU frequencies (cf.~Section \ref{subsubsec:EnvironFactorDicussion}), there is also a debate on energy saving by downscaling the CPU frequency. Considering the cubic relationship between a CPU's power and its clock frequency (cf.~Equation \ref{eqn:DVFS}), in theory, three quarters of the energy can be saved by halving the processor's clock speed, although the execution time doubles \cite{Kumar_Lu_2010}. In practice, unfortunately, blindly downscaling CPU frequency often increases energy consumption \cite{Liu_Pinto_2015}, and computation-intensive applications would particularly be less energy efficient when operating processors at lower frequencies \cite{Zhang_Fu_2011}. Such a debate is still driven by the aforementioned ``race to idle", depending on if reducing power consumption can bring overwhelming energy benefits.


\item \textbf{Workload Offloading:} In mobile Cloud computing, offloading local workloads to external resources have widely been considered effective to shorten applications' execution time and extend mobile devices' battery life, because powerful remote servers can generally offer a significant speedup for mobile applications \cite{Chun_Ihm_2011,Nabi_Mittal_2015}. However, simply offloading workloads has been proven not always to be energy efficient \cite{Vishwanath_Jalali_2015}, unless the workload is characterized by a relatively small communication-computation ratio \cite{Kumar_Lu_2010}. Correspondingly, the communication-computation ratio has frequently been employed as a trade-off indicator to help determine the right circumstances of workload offloading \cite{Miettinen_Nurminen_2010,Mtibaa_Harras_2013,Namboodiri_Ghose_2013}.

\end{itemize}

\section{Conclusions and Future Work}
\label{sec:conclusion}
The energy consumption of Cloud computing is predicted to keep growing and even quadruple the current annual consumption by 2020 \cite{Guyon_Orgerie_2015}. Thus, efficient use of computing power and energy consumption management have become crucial topics for engineering Cloud applications. With modeling as a prevalent approach to addressing energy consumption, a substantially large number of models with a high variety has emerged. This drives us to use SLR as a rigorous surveying approach to study the existing modeling efforts as evidence to build up a knowledge foundation for investigating Cloud applications' energy consumption. 

In particular, by deconstructing Cloud computing scenarios, we find that the controllable environmental components (especially client devices) and the application execution elements related to task processing and data communication have attracted most of the research attention as well as the modeling efforts. 
By identifying energy-related factors, this survey confirms computation and communication to be the existing researchers' major concerns about energy consumption of Cloud applications. Correspondingly, Task Size and Data Size have been considered to be the main workload factors, which would largely interact with CPU Clock Frequency and Network Bandwidth (and Access Point Technology used in the client devices) as main environmental factors. On the contrary, the energy consumption of data storage has attracted little attention, and few studies have intensively investigated and modeled the energy for Cloud applications' memory footprints. Such a finding indicates crucial research gaps that require further research efforts in the future.

In fact, storage policies in different cloud environments, which partly relates to the application's nature, may result in a considerably high persistence of the application's data in the Cloud storage, and in turn gives rise to energy consumption for keeping the data. Not to mentions that the degree of data distribution (for protection purposes) can also negatively affect the energy consumption of data storage. Meanwhile, given the increasing trend of in-memory Cloud computing (e.g., Apache Spark\footnote{\url{http://spark.apache.org/}}), memory has become a significant contributor to the power consumption in Cloud infrastructures \cite{Chen_Wang_2011}.

More importantly, our work has advocated divide-and-conquer to be a principle approach to studying energy consumption in the Cloud computing domain. On one hand, decomposing an energy consumption scenario can help clarify the atomic energy concerns and mitigate the complexity in the corresponding problem. On the other hand, gradually recomposing major energy concerns can facilitate iterative and incremental development of energy consumption models, in order to address the complicated trade-offs and even debates with respect to energy efficiency. Naturally, we will unfold our future work along two directions. The first direction is to gradually expand the knowledge artefact (including both factors and models) established in this survey. The second direction is to implement model-driven simulations to reveal further knowledge about the combinational factorial effects on Cloud applications' energy consumption.

\bibliography{Energy-Ref-Abbr}

\end{document}